\documentclass[acmsmall,natbib=true,screen=true]{acmart}

\usepackage{acronym}
\usepackage[skip=2pt]{caption}
\usepackage[inline]{enumitem}
\usepackage{booktabs} 
\usepackage{multirow}
\usepackage{array}
\usepackage{pifont}
\usepackage{xspace}
\usepackage{subcaption}
\usepackage{makecell}
\usepackage{CJKutf8}
\usepackage{pifont}
\usepackage{tcolorbox}

\acrodef{EEwRJ}{effectiveness evaluation without relevance judgments}
\acrodef{RAG}{retrieval-augmented generation}
\acrodef{LLM}{large language model}
\acrodef{QPP}{query performance prediction}
\acrodef{IR}{information retrieval}
\acrodef{NLP}{natural language processing}
\acrodef{PEFT}{parameter-efficient fine-tuning}
\acrodef{ICL}{in-context learning}
\acrodef{LoRA}{low-rank adaptation}

\acrodef{RR}{reciprocal rank}
\acrodef{AP}{Average Precision}
\acrodef{nDCG}{normalized discounted cumulative gain}

\acrodef{HSD}{Tukey's honestly significant difference}
\acrodef{ANOVA}{analysis of variance}

\acrodef{UEF}{utility estimation framework}
\acrodef{QPP-PRP}{pairwise rank preference-based QPP}
\acrodef{M-QPPF}{multi-task query performance prediction framework}
\acrodef{WRIG}{weighted relative information gain-based model}
\acrodef{NLP}{natural language processing}
\acrodef{CIS}{conversational information seeking}
\acrodef{ANCE}{Approximate nearest neighbor Negative Contrastive Estimation}

\acrodef{RL}{reinforcement learning}

\AtBeginDocument{%
  \providecommand\BibTeX{{%
    \normalfont B\kern-0.5em{\scshape i\kern-0.25em b}\kern-0.8em\TeX}}}

\newcommand{\negar}[1]{\textcolor{pink}{\textbf{[#1]}}}

\newcommand{\ourmodel}{QPP-GenRE\xspace}

\makeatletter
\@ifpackagelater{acronym}{2020/01/01}
{%
}
{%

\newcommand{\Ac}[1]{\ac{#1}(\textbf{properly capitalized})}
\newcommand{\Acf}[1]{\acf{#1}(\textbf{properly capitalized})}
}%
\makeatother

\setcopyright{rightsretained}
\acmJournal{TOIS}
\acmYear{2025} 
\acmVolume{1} \acmNumber{1} \acmArticle{1} \acmMonth{1} \acmPrice{}
\acmDOI{10.1145/3736402}

\author{Chuan Meng}
\orcid{0000-0002-1434-7596}
\affiliation{
  \institution{University of Amsterdam}
  \city{Amsterdam}
  \country{The Netherlands}
}
\email{c.meng@uva.nl}

\author{Negar Arabzadeh}
\orcid{0000-0002-4411-7089}
\affiliation{
  \institution{University of Waterloo}
  \city{Waterloo}
  \country{Canada}
}
\email{narabzad@uwaterloo.ca}

\author{Arian Askari}
\orcid{0000-0003-4712-832X}
\affiliation{
  \institution{Leiden University}
  \city{Leiden}
  \country{The Netherlands}
}
\email{a.askari@liacs.leidenuniv.nl}

\author{Mohammad Aliannejadi}
\orcid{0000-0002-9447-4172}
\affiliation{
  \institution{University of Amsterdam}
  \city{Amsterdam}
  \country{The Netherlands}
}
\email{m.aliannejadi@uva.nl}

\author{Maarten de Rijke}
\orcid{0000-0002-1086-0202}
\affiliation{%
  \institution{University of Amsterdam}
  \city{Amsterdam}
  \country{The Netherlands}
}
\email{m.derijke@uva.nl}

\ccsdesc[500]{Information systems~Evaluation of retrieval results}

\keywords{Query performance prediction, Large language models, Relevance judgments, Relevance prediction, Re-ranking, Conversational search}

\begin{document}

\title[Query Performance Prediction using Relevance Judgments Generated by Large Language Models]{Query Performance Prediction using Relevance Judgments Generated by Large Language Models}

\renewcommand{\shortauthors}{Meng et al.}


\begin{abstract}
\Acf{QPP} aims to estimate the retrieval quality of a search system for a query without human relevance judgments.
Previous \ac{QPP} methods typically return a single scalar value and do not require the predicted values to approximate a specific \ac{IR} evaluation measure, leading to certain drawbacks:
\begin{enumerate*}[label=(\roman*)]
    \item a single scalar is insufficient to accurately represent different \ac{IR} evaluation measures, especially when metrics do not highly correlate, and 
    \item a single scalar limits the interpretability of \ac{QPP} methods because solely using a scalar is insufficient to explain \ac{QPP} results. 
\end{enumerate*} 
To address these issues, we propose a \ac{QPP} framework using automatically \textbf{gen}erated \textbf{re}levance judgments (\ourmodel), which decomposes \ac{QPP} into independent subtasks of predicting the relevance of each item in a ranked list to a given query.
This allows us to predict any \ac{IR} evaluation measure using the generated relevance judgments as pseudo-labels.
This also allows us to interpret predicted \ac{IR} evaluation measures, and identify, track and rectify errors in generated relevance judgments to improve \ac{QPP} quality.
We predict an item's relevance by using \textit{open-source} \acp{LLM} to ensure scientific reproducibility.

We face two main challenges:
\begin{enumerate*}[label=(\roman*)]
    \item excessive computational costs of judging an entire corpus for predicting a metric considering recall, and 
    \item limited performance in prompting open-source \acp{LLM} in a zero-/few-shot manner.
\end{enumerate*} 
To solve the challenges, we devise an approximation strategy to predict an \ac{IR} measure considering recall and propose to fine-tune open-source \acp{LLM} using human-labeled relevance judgments.
Experiments on the TREC 2019--2022 deep learning tracks and CAsT-19--20 datasets show that \ourmodel achieves state-of-the-art \ac{QPP} quality for both lexical and neural rankers. 
\end{abstract}

    

%



\maketitle

\acresetall


\section{Introduction}

\Acf{QPP}, a.k.a.~query difficulty prediction, has attracted the attention of the \acf{IR} community throughout the years~\citep{arabzadeh2025query,arabzadeh2024query1,arabzadeh2024query,ganguly2022analysis,carmel2010estimating}. 
\ac{QPP} aims to estimate the retrieval quality of a search system for a query without using human-labeled relevance judgments~\citep{faggioli2021enhanced}. 
Effective \ac{QPP} benefits various {downstream applications~\citep{faggioli2023report}, e.g., query variant selection~\citep{di2021study,scells2018query,thomas2017tasks}, 
selective query expansion~\citep{amati2004query}, 
\ac{IR} system configuration selection~\citep{deveaud2016learning,tonellotto2013efficient}, 
enriching query features for learning-to-rank~\citep{macdonald2012usefulness}
,
and
query-specific pool depth prediction~\citep{ganguly2023query} to reduce human relevance judgment costs.

\paragraph{Current limitations}
\ac{QPP} methods can be applied in various domains and scenarios~\citep{meng2025qpp,faggioli2023report}. 
We are usually concerned with the predicted retrieval quality w.r.t.\ various \ac{IR} measures across different scenarios, e.g., our emphasis might be on precision~\citep{faggioli2023geometric,faggioli2021hierarchical} for conversational search~\citep{lu2025zero,abbasiantaeb2025improving} and on recall for legal search~\citep{tomlinson2007overview}. 
However, existing \ac{QPP} approaches typically predict only a single real-valued score that indicates the retrieval quality for a query~\citep{ganguly2022analysis} and do not require the predicted score to approximate a specific \ac{IR} evaluation measure~\citep{arabzadeh2023noisy,singh2023unsupervised,tao2014query,shtok2012predicting,zhou2007query}. 

These properties results in two key limitations:
\begin{enumerate*}[label=(\roman*)]
    \item While predicted performance scores have been shown to correlate with some \ac{IR} evaluation metrics~\citep{datta2022pointwise,ganguly2022analysis}, relying on a single value to represent different \ac{IR} evaluation measures leads to a ``one size fits all'' approach, which is problematic because the literature shows that some \ac{IR} metrics do not correlate well and the agreement varies across scenarios and queries~\citep{gupta2019correlation,jones2015features}.
    Although some studies train regression-based \ac{QPP} models to predict a specific \ac{IR} evaluation measure~\citep{khodabakhsh2023learning,chen2022groupwise,datta2022deep,arabzadeh2021bert,hashemi2019performance}, they require training separate models to predict different measures, leading to lots of storage and running costs.
    \item A single-score prediction limits the interpretability of \ac{QPP}. It is insufficient to explain \ac{QPP} outputs or to analyze and fix inaccurate \ac{QPP} results based solely on a single score.
    We argue that more in-depth and interpretable insights into \ac{QPP} outputs are required. 
\end{enumerate*} 

\paragraph{A novel \ac{QPP} framework}
We propose a \ac{QPP} framework using automatically \textbf{gen}erated \textbf{re}levance judgments (\ourmodel), in which
we decompose \ac{QPP} into independent subtasks of automatically predicting the relevance of each item in a ranked list for a given query.
\ourmodel comes with various advantages:
\begin{enumerate*}[label=(\roman*)]
    \item It allows us to directly predict any desired \ac{IR} evaluation measure at no additional cost, using automatically generated relevance judgments as pseudo-labels.
Compared to most existing QPP methods that only outputting a single scalar value as an indicator of a ranker's overall performance, our method provides a multi-dimensional assessment of system effectiveness by enabling the calculation of various metrics from the same set of relevance judgments. 
Leveraging predicted relevance judgments, our method allows the computation of metrics such as nDCG@k, Precision@k, \ac{RR}, and so on. 
This flexibility is particularly advantageous because it allows us to use \ac{QPP} to predict retrieval quality in terms of a specific evaluation metric that is prioritized in different scenarios.
For example, we use predicted relevance judgments to calculate precision-oriented metrics in conversational search, while focusing on recall-oriented metrics for tasks like legal search.
%
    \item The generated relevance judgments provide an explanation beyond simply gauging how difficult or easy a query is by offering information about why the query is predicted as being difficult or easy; moreover, we can translate the ``\ac{QPP} errors'' into easily observable ``relevance judgment errors,'' e.g., false positives or negatives, informing potential ways of improving \ac{QPP} quality by fixing observed relevance judgment errors. 
\end{enumerate*}



\paragraph{Integrating \ourmodel with LLM-labeled relevance judgments}
\ourmodel can be integrated with various approaches for judging relevance.
The success of \ourmodel depends fundamentally on the accuracy of relevance judgment predictions.
Therefore, it is crucial to equip \ourmodel with an approach capable of accurately generating relevance judgments.
Recently, numerous studies~\citep{takehi2024llm,upadhyay2024large,ma2024leveraging,thomas2023large,faggioli2023perspectives,macavaney2023one,gilardi2023chatgpt} have shown the potential effectiveness of using \acp{LLM} to generate relevance judgments.
Therefore, it is natural to explore equipping \ourmodel with \acp{LLM} for judging relevance.
However, those studies have certain limitations: 
several authors have prompted commercial \acp{LLM} (e.g., ChatGPT, GPT-3.5/4, GPT-4o) to generate relevance judgments~\citep[e.g.,][]{chen2024ai,upadhyay2024umbrela,upadhyay2024llms,zhang2024large,ma2024leveraging,thomas2023large,faggioli2023perspectives}; commercial \acp{LLM} come with limitations like non-reproducibility, non-deterministic outputs and potential data leakage between pretraining and evaluation data, impeding their value in scientific research~\citep{pradeep2023rankzephyr,zhang2023rank,pradeep2023rankvicuna}.
Although \citet{macavaney2023one} prompt small-scale open-source language models (e.g., Flan-T5~\citep{chung2022scaling} with 3B parameters) for generating relevance judgments, they focus on a setting wherein the model is already given one relevant item for each query, which does not apply to \ac{QPP} as we typically do not know any relevant item for a query in advance.
%
%
In this paper, we focus on the use of \textit{open-source} \acp{LLM} for generating relevance judgments in a realistic setting where we lack prior knowledge of any relevant items for a query.
There are only few studies~\citep{zhang2024large,salemi2024evaluating,upadhyay2024llms,khramtsova2024leveraging} attempting to prompt open-source \acp{LLM} in this setting.

%

%

\paragraph{Challenges}
We face two challenges when using \ourmodel for \ac{QPP}:
\begin{enumerate*}[label=(\roman*)]
\item Predicting \ac{IR} metrics that not only consider precision but also take recall into account, ideally, entails identifying all relevant items in the entire corpus for a query; however, using an \ac{LLM} to judge the entire corpus per query is impractical due to the significant computational overhead;
\item our experiments reveal that directly prompting open-source \acp{LLM} in a zero-/few-shot manner yields limited effectiveness in predicting relevance, resulting in limited \ac{QPP} quality; this aligns with recent findings indicating limited success in prompting open-source \acp{LLM} for specific tasks~\citep{qin2023large}.
Also, incorporating in-context learning examples in few-shot prompting leads to high inference costs~\citep{liu2022few}.
\end{enumerate*}

\paragraph{Solutions}
To address the challenges listed above, 
\begin{enumerate*}[label=(\roman*)]
\item we devise an approximation strategy to predict \ac{IR} measures considering recall by only judging a few items in the ranked list for a query and using them to estimate the metric, hence avoiding the cost of traversing the entire corpus to identify all relevant items for a query; the approximation strategy also enables us to investigate the impact of various judging depths in the ranked list on \ac{QPP} quality; and 
\item we enhance an open-source \ac{LLM}'s ability to generate relevance judgments by training it with \ac{PEFT}~\citep{dettmers2023qlora} on human-labeled relevance judgments; unlike previous supervised \ac{QPP} methods that need to train separate models for predicting different \ac{IR} evaluation measures, training \acp{LLM} to judge relevance is agnostic to a specific \ac{IR} metric.
\end{enumerate*}

\paragraph{Experiments}
%
Experiments on datasets from the TREC 2019--2022 deep learning (TREC-DL) tracks~\citep{craswell2022,2021craswell,craswell2020,craswell2019} show that \ourmodel achieves state-of-the-art \ac{QPP} quality in estimating the retrieval quality of a lexical ranker (BM25) and two neural rankers, ANCE~\citep{xiong2021approximate} and TAS-B~\citep{hofstatter2021efficiently}, in terms of RR@10, a precision-oriented IR metric, and nDCG@10, an \ac{IR} metric considering recall.
See Sections~\ref{subsec:precision} and \ref{subsec:recall}.   

We also find that using \acp{LLM} to directly model \ac{QPP}, i.e., asking \acp{LLM} to directly generate values of \ac{IR} evaluation metrics, performs much worse than \ourmodel. 
This finding reveals that \ourmodel is a more effective way of modeling \ac{QPP} using \acp{LLM}.
Furthermore, our experiments demonstrate the effectiveness of our devised approximation strategy in nDCG@10: \ourmodel achieves state-of-the-art \ac{QPP} quality at the shallow judging depth 10, and \ourmodel's \ac{QPP} quality reaches saturation when it further judges up to 100--200 retrieved items in a ranked list.
See Section~\ref{sec:depth}.

Moreover, we conduct an in-depth analysis to investigate the impact of fine-tuning and the choice of \acp{LLM} on the quality of generated relevance judgments and \ac{QPP}.
We consider two families of \acp{LLM}, Llama and Mistra, with sizes ranging from 1B to 70B, under both few-shot and fine-tuned settings.
We find that fine-tuning markedly improves the quality of relevance judgment generation and \ac{QPP} for all \acp{LLM}.
In particular, a fine-tuned 3B model (Llama-3.2-3B-Instruct) provides the best trade-off between \ac{QPP} quality and computational efficiency: it not only significantly outperforms 70B few-shot models, but also achieves \ac{QPP} quality comparable to that of fine-tuned 7B and 8B models.
This suggests that, compared to few-shot prompting, fine-tuning LLMs for relevance prediction can yield higher effectiveness in both relevance prediction and QPP, even with relatively small model sizes; this, in turn, implies that fine-tuning can offer strong performance at lower inference costs.
Moreover, the performance of fine-tuned \acp{LLM} in terms of judging relevance exceeds that of a commercial \ac{LLM} (GPT-3.5)~\citep{faggioli2023perspectives}.
See Section~\ref{sec:fewshot}.


Additionally, to show \ourmodel's compatibility with other types of relevance prediction methods, we adapt a state-of-the-art pointwise \ac{LLM}-based re-ranker, RankLLaMA~\citep{ma2023fine}, into a relevance judgment generator by applying a threshold to its re-ranking scores.
Our results indicate that \ourmodel integrated with RankLLaMA achieves high \ac{QPP} quality, at the cost of tuning a proper threshold.
The high \ac{QPP} quality achieved by RankLLaMA demonstrates \ourmodel's compatibility with other types of relevance prediction methods.
See Section~\ref{sec:reranker}.

To demonstrate the generalizability of QPP-GenRE to a new domain, we conduct experiments of applying QPP-GenRE to conversational search~\citep{mo2025conversational,meng2025bridging} in a zero-shot manner. 
Specifically, we evaluate QPP-GenRE and baselines when predicting the performance of a conversational dense retriever~\citep{yu2021few} on the CAsT-19~\citep{dalton2020cast} and 20~\citep{Dalton2020CAsT2T} datasets. 
We found that QPP-GenRE consistently outperforms all baselines on both datasets, demonstrating strong generalizability.
See Section~\ref{sec:cs}.

%
We also analyze \ac{QPP} errors based on automatically generated relevance judgments, and provide a case study for a specific example, demonstrating \ourmodel's interpretability.
See Section~\ref{sec:interpretability}.

Finally, our computational cost analysis shows that \ourmodel shows lower latency than some supervised \ac{QPP} baselines when predicting multiple measures because multiple measures can be derived from the same set of relevance judgments.
Although \ourmodel shows higher latency than other \ac{QPP} baselines when predicting only one metric, \ourmodel’s latency is still 20 times smaller than the state-of-the-art GPT-4-based listwise re-ranker~\citep{sun2023chatgpt}.
To further enhance the efficiency of \ourmodel. 
We have proposed a \textit{relevance judgment caching mechanism}. 
Our experimental results show that the mechanism can reduce LLM calls for relevance prediction by about 30\%. Specifically, this mechanism reuses previously predicted relevance judgments for the same query when predicting the performance of new rankers. 
As a result, this mechanism helps conserve computational resources by avoiding recompute relevance judgments that are shared among multiple rankers.
See Section~\ref{sec:cost}.

\paragraph{Application scenarios}
Given \ourmodel's high \ac{QPP} quality and interpretability, it is well-suited for some knowledge-intensive professional search scenarios, where accurate \ac{QPP} is prioritized, interpretable \ac{QPP} results are preferred, and users may have a higher tolerance level for latency than users in web search.
Plus, \ourmodel can be used to analyze how well a search system performs in offline settings~\citep{faggioli2023geometric}, where latency is not necessarily an issue.

One might argue, if \ourmodel needs to be integrated with an \ac{LLM} to predict ranking quality, why not directly use the \ac{LLM} for re-ranking?
However, we reveal that \ourmodel integrated with LLaMA-7B already achieves high \ac{QPP} quality and remains significantly more efficient than costly state-of-the-art \ac{LLM}-based re-rankers (e.g., the GPT-4-based listwise re-ranker~\citep{sun2023chatgpt}).
Calling those expensive \ac{LLM}-based re-rankers is often unnecessary, as many initial rankings are good enough and either do not require re-ranking or only need very shallow re-ranking depths~\citep{meng2024ranked}.
Therefore, sufficiently accurate \ac{QPP} for initial rankings is needed to guide the decision on whether to use the expensive re-ranker, or to determine the optimal re-ranking depth that does not waste computational resources.
Given \ourmodel's substantial improvements in \ac{QPP} quality over previous \ac{QPP} methods and significantly lower latency compared to those expensive re-rankers, it is valuable to make \ourmodel work with state-of-the-art, yet much more costly, \ac{LLM}-based re-rankers~\citep{sun2023chatgpt} to achieve a better balance between effectiveness and efficiency in re-ranking.

Another advantage of \ourmodel, which makes it applicable to real-world scenarios, is that unlike traditional approaches that depend heavily on the specific properties of rankers, our method is ranker-agnostic. 
E.g., conventional baselines often rely on score distributions tied to the type of their rankers, making their predictions inherently ranker-dependent. 
Previous study has demonstrated that the effectiveness of such score-based QPP methods varies across different rankers due to the differences in score distributions produced by each ranker \citep{faggioli2023query}. 
In contrast, \ourmodel operates on individual query--document pairs and evaluating them independently of a ranker. 
This eliminates the dependency on specific ranker characteristics and score distributions, ensuring that our framework can be applied generally across various retrieval settings. 
Furthermore, \ourmodel can leverage the reusability of predicted relevance judgments.
Since each query--document pair is judged only once, our method allows for predicting the performance of multiple rankers effectively due to their potential overlaps in their top ranked documents. 
This implies that \ourmodel can become more efficient over time as it is used in practice.


\paragraph{Reproducibility}
To facilitate future research, we release our data, scripts for fine-tuning/inference, sampled demonstration examples for few-shot prompting, and fine-tuned checkpoints of various \acp{LLM} at \url{https://github.com/ChuanMeng/QPP-GenRE}.

\paragraph{Contributions}
Our main contributions are as follows:
\begin{itemize}
\item We propose a novel \ac{QPP} framework using automatically generated relevance judgments (\ourmodel), which decomposes \ac{QPP} into independent subtasks of predicting the relevance of each item in a ranked list to the query, and predicts different \ac{IR} evaluation measures based on the relevance predictions.
\item We devise an approximation strategy to predict \ac{IR} measures that account for both precision and recall, avoiding the cost of traversing the entire corpus to identify all relevant items for a query.
\item We fine-tune leading \textit{open-source} \acp{LLM} from the Llama and Mistral families, covering a range of model sizes, for the task of automatically generating relevance judgments.
Our results show that fine-tuning much smaller \acp{LLM} for relevance judgment prediction can yield more effective relevance prediction and \ac{QPP} than few-shot prompting with much larger models.
\item We conduct experiments on four datasets, showing that \ourmodel outperforms the state-of-the-art \ac{QPP} baselines on the TREC-DL 19--22 datasets in predicting RR@10 and nDCG@10 in terms of Pearson's $\rho$ and Kendall's $\tau$. 
\end{itemize}

\section{Related Work}
Our work is relevant to four strands of research: \acf{QPP} (Section~\ref{sec:qpp}), zero/few-shot prompting and \acf{PEFT} for \acp{LLM} (Section~\ref{sec:peft}), \acp{LLM} for generating relevance judgments (Section~\ref{sec:rj}), and \acp{LLM} for re-ranking (Section~\ref{sec:reranking}).

\subsection{Query performance prediction}
\label{sec:qpp}
\Acf{QPP} has attracted lots of attention in the \ac{IR} and NLP community and has been widely studied in ad-hoc search~\citep{singh2023unsupervised,faggioli2023towards,faggioli2023query,datta2022pointwise}, conversational search~\citep{faggioli2023spatial,faggioli2023geometric,abbasiantaeb2024llm,meng2023query,meng2023Performance,sun2021conversations,meng2024dc,meng2023system}, question answering~\citep{samadi2023performance, hashemi2019performance}, and image retrieval~\citep{poesina2023iqpp}.
This paper focuses on \ac{QPP} for ad-hoc search.

Typically, \ac{QPP} methods are divided into two categories: pre- and post-retrieval methods~\citep{carmel2010estimating}.
The former predicts the difficulty of a given query
by using features of the query and corpus, while the latter further uses features of a ranked list returned by a ranker for the query~\citep{carmel2010estimating}. 
This paper focuses on post-retrieval \ac{QPP} methods.

A large number of unsupervised and supervised post-retrieval \ac{QPP} methods have been proposed \citep{carmel2010estimating} for predicting the performance of lexical rankers, such as query likelihood~\citep{lafferty2001document} and BM25~\citep{robertson1995okapi}.
Unsupervised \ac{QPP} methods can be classified into clarity-based~\citep{cronen2002predicting}, robustness-based~\citep{zhou2007query,zhou2006ranking,aslam2007query}, coherence-based~\citep{arabzadeh2021query,diaz2007performance}, and score-based~\citep{tao2014query,shtok2012predicting,cummins2011improved,perez2010standard,zhou2007query}.
%
More recently, a set of supervised \ac{QPP} methods have been proposed \cite{arabzadeh2021bert,hashemi2019performance,zamani2018neural,datta2022deep,datta2022relative,chen2022groupwise,khodabakhsh2023learning}.
NeuralQPP~\citep{zamani2018neural} and Deep-QPP~\citep{datta2022deep} are optimized from scratch.
NQA-QPP~\citep{hashemi2019performance} and BERT-QPP~\citep{arabzadeh2021bert} fine-tune BERT~\citep{devlin2019bert} to improve \ac{QPP} effectiveness.
Further, \citet{datta2022pointwise} propose qppBERT-PL, which considers list-wise-document information, while \citet{chen2022groupwise} propose BERT-groupwise-QPP that considers both cross-query and cross-document information.
\citet{khodabakhsh2023learning} propose a \ac{M-QPPF}, learning document ranking and \ac{QPP} simultaneously.

Post-retrieval \ac{QPP} methods designed for lexical rankers struggle to predict the retrieval quality of neural rankers~\citep{faggioli2023query,hashemi2019performance}, motivating several new unsupervised post-retrieval \ac{QPP} methods designed for neural rankers.
\citet{datta2022relative} propose a \ac{WRIG}, which assesses a neural ranker for a given query by considering the relative difference of predicted performance between the given query and its variants; 
\citet{zendel2023entropy} assess a neural re-ranker by measuring the entropy of scores returned by it; 
\citet{faggioli2023towards} propose neural-ranker-specific ways of calculating regularization terms used by unsupervised post-retrieval \ac{QPP} methods; 
\citet{vlachou2023coherence} propose an unsupervised coherence-based \ac{QPP} method that employs neural embedding representations to assess dense retrievers; and
\citet{singh2023unsupervised} propose \ac{QPP-PRP} for predicting the performance of a neural ranker by measuring the degree to which a pairwise neural re-ranker (e.g., DuoT5~\citep{spradeep2021expando}) agrees with the ranked list returned by the neural ranker.  

We present a novel \ac{QPP} perspective: we start by automatically generating relevance judgments for a ranked list for a query and then proceed to predict  \ac{IR} evaluation measures for the ranked list. 
To the best of our knowledge, no prior work addresses \ac{QPP} from this perspective. 

Unlike regression-based \ac{QPP} models~\citep{khodabakhsh2023learning,chen2022groupwise,datta2022deep,arabzadeh2021bert,hashemi2019performance}, which require training separate models to predict different \ac{IR} evaluation measures, the training of \acp{LLM} for judging relevance in the \ourmodel method that we propose is agnostic to a specific \ac{IR} evaluation measure, and different measures can be derived from the same set of generated relevance judgments. 

We also differ from qppBERT-PL~\citep{datta2022pointwise}, which first predicts the number of relevant items for each chunk in a ranked list and then aggregates those numbers into a general \ac{QPP} score.
However, qppBERT-PL's output is still presented as a single scalar, which is insufficient to accurately represent different evaluation measures; also, it is infeasible to predict arbitrary IR measures only using the number of relevant items in a ranked list.

The work closest to \ourmodel, which is still different, is \ac{QPP} using \ac{EEwRJ}~\citep{mizzaro2018query}.
The goal of \ac{EEwRJ} methods is to predict search system effectiveness in a TREC-like environment.
E.g., a method proposed by \citet{soboroff2001ranking} randomly samples items from a pool for a query and treats these items as relevant; the intuition is that if an item is ranked highly by many search systems, it is likely to be pooled and therefore considered relevant.
\citet{mizzaro2018query} explore applying \ac{QPP} \ac{EEwRJ}~\citep{mizzaro2018query} methods to \ac{QPP}.
However, \ac{QPP} using \ac{EEwRJ} suffers from two limitations:
\begin{enumerate*}[label=(\roman*)]
\item \ac{EEwRJ} requires obtaining ranked lists returned by all search systems in a given TREC edition to predict the difficulty of a query, and
\item \ac{EEwRJ} encounters normalization challenges when predicting the ranking quality for a ranked list returned by a specific search system~\citep{mizzaro2018query}.
\end{enumerate*} 
\ourmodel does not face these limitations.

\subsection{Zero/few-shot prompting and parameter-efficient fine-tuning for LLMs}
\label{sec:peft}

While fine-tuning pre-trained language models has given rise to many state-of-the-art results~\citep{devlin2019bert}, fully fine-tuning \acp{LLM} for a specific task on consumer-level hardware is typically infeasible~\citep{zhu2023large} because of the large number of parameters of \acp{LLM}. 
As a result, there are three prevailing ways to adapt \acp{LLM} to a specific task: zero-shot prompting, few-shot prompting~\citep{askari2025self,askari2024generative}, a.k.a.\ \ac{ICL}~\citep{dong2022survey,brown2020language}, and \acf{PEFT}~\citep{dettmers2023qlora,liu2022few,hu2021lora}.
%

There is limited success in only prompting open-source \acp{LLM} for certain tasks~\citep{qin2023large}.
Zero-shot prompting instructs an \ac{LLM} to perform a specific task by inputting a text instruction.
To get a promising result, zero-shot prompting is usually based on instruction-tuned \acp{LLM}~\citep{zhang2023instruction,qin2023large}, such as Flan-T5~\citep{chung2022scaling}, Flan-UL2~\citep{tay2022ul2}.
However, \citet{sun2023evaluating} show that the performance of zero-shot prompting degrades considerably if an LLM is fed an instruction that was not observable during its training.
\ac{ICL} inputs a few input-target pairs (a.k.a. demonstrations) to an \ac{LLM}, which would make an \ac{LLM} learn from analogy~\citep{dong2022survey} without updating its parameters.
However, \ac{ICL} has a high computational cost because it needs to feed input-target pairs to an \ac{LLM} for each prediction; also, \ac{ICL} requires substantial manual prompt engineering because an \ac{LLM}'s performance~\citep{liu2022few} is sensitive to the formatting of the prompt (e.g., the wording and the order of input-target pairs).  

\ac{PEFT} can solve the above limitations; it aims to adapt an \ac{LLM} to a specific task by training only a small fraction of its parameters.
\Ac{LoRA}, a widely-used \ac{PEFT} method~\citep{gema2023parameter,liu2023goat,zhang2023multi,santilli2023camoscio}, has been shown to achieve comparable performance to full-model fine-tuning~\citep{dettmers2023qlora,lu2023empirical}; \ac{LoRA} adds learnable low-rank adapters to each network layer of an \ac{LLM}~\citep{hu2021lora} while all original parameters of the \ac{LLM} are frozen.  
QLoRA~\citep{dettmers2023qlora} further reduces the memory usage of \ac{LoRA} without sacrificing performance; QLoRA first quantizes an \ac{LLM} model to 4-bits before adding and optimizing low-rank adapters. 
Our work explores the possibility of training open-source \acp{LLM} with QLoRA to generate relevance judgments.

\subsection{LLMs for generating relevance judgments}
\label{sec:rj}
Automatically generating relevance judgments is a long-standing goal in \ac{IR} that has been studied for multiple decades~\citep{makary2017using,makary2016towards,ravana2015ranking,nuray2006automatic,nuray2003automatic,soboroff2001ranking}.
Recent studies have demonstrated promising results of using \acp{LLM} for the automatic generation of relevance judgments~\citep{thomas2023large,faggioli2023perspectives}.
In this paper we focus on studies into generating relevance judgments with discrete classes (e.g., ``Relevant'' or ``Irrelevant''), instead of generating continuous relevance labels in real numbers~\citep{yan2024consolidating}.
We discuss related studies into \ac{LLM}-based automatic generation of relevance judgments from two dimensions: (i)~how \acp{LLM} are used to generate relevance judgments, and (ii)~their applications. 

Recent studies have explored prompting commercial \acp{LLM} (e.g., GPT-3.5 and GPT 4) or open-source \acp{LLM} in zero- or few-shot manners. 
Specifically, \citet{faggioli2023perspectives} use zero- and few-shot prompting to instruct GPT-3.5 to predict the relevance of an item to a query.  
\citet{thomas2023large} instruct GPT-4 by zero-shot prompting, and add to the prompt a detailed query description and consider chain-of-thought~\citep{wei2022chain}.
%
%
\citet{ma2024leveraging} instruct GPT-3.5 to generate relevance judgments for a domain-specific scenario, i.e., legal case retrieval~\citep{ma2021lecard}; they use prompts specifically designed for this scenario.
More recently, \citet{upadhyay2024umbrela} prompt GPT-4o in a zero-shot manner.
Besides using commercial \acp{LLM}, only few  studies~\citep{zhang2024large,salemi2024evaluating,upadhyay2024llms,khramtsova2024leveraging} explore prompting open-source \acp{LLM} to generate relevance judgments.
E.g., \citet{khramtsova2024leveraging}, \citet{upadhyay2024llms} and \citet{salemi2024evaluating} prompt Flan-T5~\citep{chung2022scaling}, Vicuña-7B~\citep{zheng2024judging} and Mistral~\citep{jiang2023mistral}, respectively, in either zero-shot or few-shot manners.
\citet{macavaney2023one} focus on a special scenario where a relevant item for a given query is already known and use Flan-T5~\citep{chung2022scaling} to estimate the relevance of another item to the query given the known relevant item.

Recent studies have explored using \ac{LLM}-generated relevance judgments to benefit \begin{enumerate*}[label=(\roman*)]
\item search system evaluation~\citep{upadhyay2024llms,thomas2023large,macavaney2023one,faggioli2023perspectives},
\item ranker selection~\citep{khramtsova2024leveraging},
\item item selection and retrieval quality evaluation in \ac{RAG}~\citep{zhang2024large,salemi2024evaluating} and
\item retriever fine-tuning~\citep{ma2024leveraging}.
\end{enumerate*} 
Concerning (i), recent studies~\citep{upadhyay2024llms,abbasiantaeb2024can,faggioli2023perspectives,thomas2023large,macavaney2023one} explore evaluating search systems either entirely using \ac{LLM}-generated relevance judgments or partially using \ac{LLM}-generated relevance judgments (a.k.a. filling holes).
They have demonstrated a high correlation between search system rankings based on \ac{LLM}- and human-labelled relevance judgments. 
As to (ii), given a pool of dense retrievers, \citet{khramtsova2024leveraging} select a suitable one for a target corpus by estimating their performance using \ac{LLM}-generated queries and relevance judgments specific to the target corpus.
For (iii), for item selection, \citet{zhang2024large} prompt \acp{LLM} to generate relevance judgments for retrieved candidate items in \ac{RAG}; the items that are predicted as ``relevant'' are used for text generation.
\citet{zhang2024large} observe that items selected via relevance prediction resulted in sub-optimal text generation quality.
For retrieval quality evaluation, \citet{salemi2024evaluating} generate relevance judgments for retrieved candidate items and aggregate those judgments into a score.
However, \citet{salemi2024evaluating} found that the aggregated score based on the \ac{LLM}-generated relevance judgments achieves a low correlation with the text generation quality of \ac{RAG}.
Concerning (iv), \citet{ma2024leveraging} fine-tune a legal case retriever on a training set augmented with \ac{LLM}-generated relevance judgments.
They show that fine-tuning a legal case retriever using the generated relevance judgments results in enhanced performance.

Our work differs from the studies mentioned above:
\begin{enumerate*}[label=(\roman*)]
\item we explore the possibility of \textit{fine-tuning} open-source \acp{LLM} for generating relevance judgments; unlike \citet{macavaney2023one}, we focus on a more practical scenario wherein no relevant item is known in advance for each query; and 
\item we focus on \ac{QPP} and predict the ranking quality of a ranked list for a query using \ac{LLM}-generated relevance judgments, which previous studies have not explored.
\end{enumerate*}

\subsection{LLMs for re-ranking}
\label{sec:reranking}

Recent studies on using \acp{LLM} for re-ranking have witnessed remarkable progress~\citep{meng2024ranked,askari2023expand,zhuang2023beyond,bommasani2023holistic,ma2023fine,ma2023zero,zhuang2023open,drozdov2023parade,sachan2022improving,zhang2023rank,pradeep2023rankzephyr,tang2023found,pradeep2023rankvicuna,ma2023zero,sun2023chatgpt,zhuang2023setwise,hou2023large}.
There are four paradigms of \ac{LLM}-based re-ranking: pointwise, pairwise, listwise, and setwise~\citep{zhuang2023setwise}. 
Given a query, pointwise re-rankers produce a relevance score for each item independently, and the final ranking is formed by sorting items by relevance score~\citep{ma2023fine,drozdov2023parade,zhuang2023beyond,sachan2022improving}.
%
The pairwise paradigm~\citep{qin2023large} eliminates the need for computing relevance scores; given a query and a pair of items, a pairwise re-ranker estimates whether one item is more relevant than the other for the query.
Listwise re-rankers~\citep{zhang2023rank,pradeep2023rankzephyr,tang2023found,pradeep2023rankvicuna,ma2023zero,sun2023chatgpt} frame re-ranking as a pure generation task and directly output the reordered ranked list given a query and a ranked list return by first-stage retriever~\citep{zhang2023rank,pradeep2023rankzephyr,tang2023found,pradeep2023rankvicuna,ma2023zero,sun2023chatgpt}.
Given the low efficiency of pairwise (multiple inference passes) and listwise (multiple decoding steps) re-rankers, 
the setwise paradigm~\citep{zhuang2023setwise} is meant to improve the efficiency while retaining re-ranking effectiveness.
Given a query and set of items, an \ac{LLM} is asked which item is the most relevant one to the query; these items are reordered according to the \ac{LLM}'s output logits of each item being chosen as the most relevant item to the query, which only requires one decoding step of an \ac{LLM}. 

Our work differs from this line of research because we generate explicit relevance judgments with discrete classes (e.g., ``Relevant'' or ``Irrelevant''), whereas studies into \acp{LLM} for re-ranking aim to predict the relevance order of items.
However, using \acp{LLM} for generating relevance judgments and for re-ranking are intrinsically the same task: relevance prediction.
Thus, an \ac{LLM}-based re-ranker has the potential to serve as a relevance judgment generator.

Our \emph{main contribution} in this paper is the introduction of \ourmodel, a novel \ac{QPP} framework, which, in theory, can be integrated with various relevance prediction approaches.
To demonstrate the compatibility of \ourmodel with various relevance prediction approaches, we adapt a state-of-the-art pointwise \ac{LLM}-based re-ranker, RankLLaMA~\citep{ma2023fine}, into a relevance judgment generator by applying a threshold for its re-ranking scores; we then integrate \ourmodel with this adapted RankLLaMA.
%
It is important to note that exploring the use of other types of \ac{LLM}-based re-rankers (e.g., pairwise and listwise) as relevance judgment generators falls outside the scope of this paper.

\section{Task definition}
In this paper, we focus on post-retrieval \ac{QPP}~\citep{carmel2010estimating}.
 Generally, a post-retrieval \ac{QPP} method $\psi$ aims to estimate the retrieval quality of a ranked list \smash{$L=[d_1,, \dots, d_i, \dots, d_{|L|}]$} with $|L|$ retrieved items induced by a ranker $M$ over a corpus $C$ in response to query $q$ without human-labeled relevance judgments, formally:
\begin{equation}
 p=\psi(q,L,C)\in \mathbb{R},
 \label{define:qpp}
\end{equation}
where $p$ indicates the predicted retrieval quality of the ranker $M$ in response to the query $q$; typically, $p$ is expected to be correlated with an \ac{IR} evaluation measure, such as \ac{RR}.


\begin{figure}[t]
\centering
    \begin{subfigure}{0.48\textwidth}
        \includegraphics[width=\textwidth]{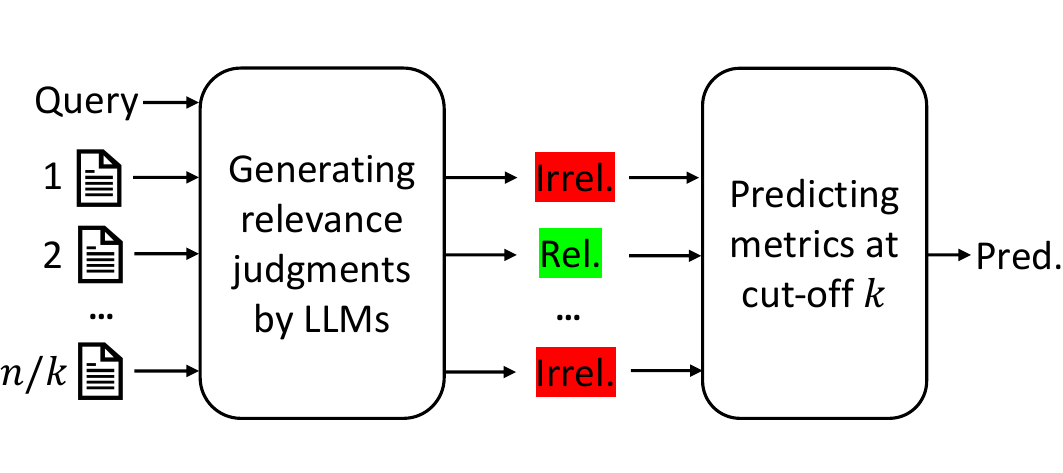} 
        \vspace{-6mm}
        \caption{Predicting a precision-based metric. In this case, we predicts the top-$n$ items in the given ranked list, $n$ is equal to $k$, which is the cut-off point for a precision-oriented metric, such as \acf{RR}. 
        E.g., the predicted $RR@k$ for the ranked list in this figure is 0.5 as the second item is predicted as relevant.}
        \label{fig:precision}
    \end{subfigure}
    \hfill
    \begin{subfigure}{0.48\textwidth}
        \includegraphics[width=\textwidth]{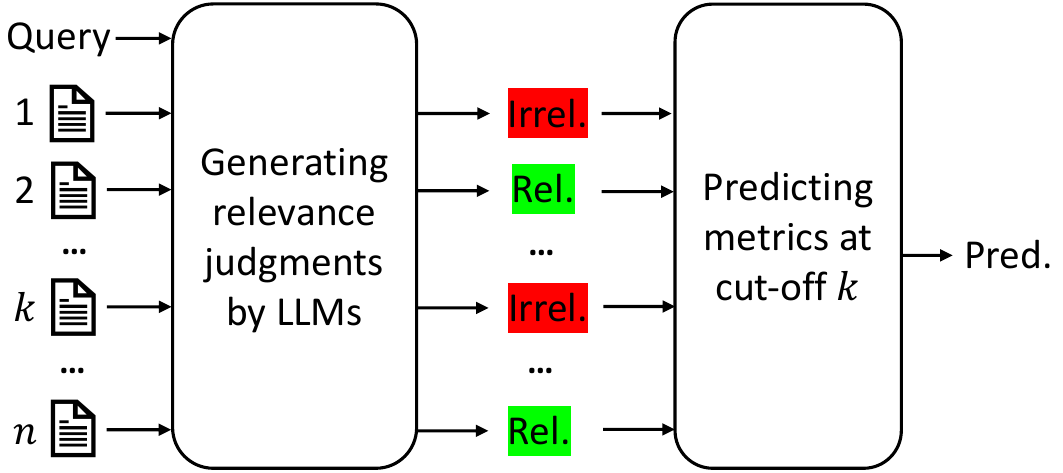}  
        \vspace{-6mm}
        \caption{Predicting a metric considering recall using our proposed \textit{approximation strategy}. 
        It only judges the top-$n$ ($n$ $\ll$ the total number of items in the corpus)} items in the given ranked list and uses the items predicted as relevant to approximate all relevant items in the corpus.
        \label{fig:recall}
    \end{subfigure}
    \caption{
    The framework of \ourmodel.
    }
    \label{fig:model-overview}
\end{figure}

\section{Method}
\label{03-model}
\subsection{Overview of \ourmodel}
We propose \ourmodel, which consists of two steps: 
\begin{enumerate*}[label=(\roman*)]
\item generating relevance judgments using \acp{LLM}, and 
\item predicting \ac{IR} evaluation measures.
\end{enumerate*} 
See Fig.~\ref{fig:model-overview}.
In (i), we employ an \ac{LLM} to generate relevance judgments for the top-$n$ retrieved items in the ranked list for a given query; to improve \acp{LLM}'s effectiveness in generating relevance judgments, we fine-tune an \ac{LLM} with \ac{PEFT} using human-labeled relevance judgments.
In (ii), we regard the generated relevance judgments as pseudo labels to calculate different \ac{IR} evaluation measures.

\subsection{Generating relevance judgments using LLMs}
\subsubsection{Inference}
Given the ranked list $L=[d_1, \dots, d_i, \dots,d_{|L|}]$ with $|L|$ items returned by a ranker $M$ for a query $q$, an \ac{LLM} is employed to automatically predict the relevance of each item in the top-$n$ positions of the ranked list $L$ to the query $q$, formally:
\begin{equation}
\hat{r}_i=\mathrm{LLM}(\mathrm{prompt}(q,d_i)),
\label{define:relevance_prediction}
\end{equation}
where $\mathrm{prompt}(\cdot, \cdot)$ is a prompt to instruct an \ac{LLM} on the task of automatic generation of relevance judgments, as illustrated in Figure~\ref{fig:prompt}.
We follow the design proposed by \citet{faggioli2023perspectives} to create a prompt that explicitly instructs the \ac{LLM} to output either ``Relevant'' or ``Irrelevant''.
In our preliminary experiments, we also tested the prompt from \citet{sun2023chatgpt}, which asks the \ac{LLM} the question, ``Does the passage answer the query?'' and expects a response of either ``Yes'' or ``No.'' However, we found that this alternative prompt produced inferior results compared to our chosen design.
$\hat{r}_i$ is a predicted relevance value for the item $d_i$ at rank $i$.
$\hat{r}_i\in\{1,0\}$ , where ``1'' indicates relevant and ``0'' irrelevant. 
We leave the prediction of multi-graded labels as future work. 
After automatically judging the top-$n$ items in the ranked list $L$, we get a list of generated relevant judgments \smash{$\hat{\mathcal{R}}_{L_{1:n}}=[\hat{r}_1, \dots, \hat{r}_i, \dots, \hat{r}_n]$}, where $\hat{r}_i$ is the predicted relevance value for $d_i$ in $L$.

\subsubsection{\Acf{PEFT}}
To further improve an \ac{LLM}'s effectiveness in generating relevance judgments, we use human-labeled relevance judgments to train an \ac{LLM} with an effective \ac{PEFT} method, QLoRA~\citep{dettmers2023qlora}.
Specifically, we first quantize an \ac{LLM} model to 4-bit, add learnable low-rank adapters to each network layer of the \ac{LLM}, and then optimize low-rank adapters. 
Formally, given the query $q$ and an item $d_i$ in the ranked list $L$, we optimize the \ac{LLM} to generate the human-labeled relevance value $r_i$ for the item $d_i$:
\begin{equation}
\label{loss}
\begin{split}
\mathcal{L}(\theta_{LoRA})=
-\frac{1}{M}\sum_{i=1}^{M}
\log P(r_i \mid \mathrm{prompt}(q,d_i)),
\end{split}
\end{equation}
where $\theta_{LoRA}$ stands for learnable low-rank adapters added to the \ac{LLM}; $M$ is the number of training examples.
See Section~\ref{implementation} for more details.

\subsection{Predicting IR evaluation measures}
\subsubsection{Predicting precision-oriented measures}
We compute a preci\-sion-oriented measure based on \ac{LLM} generated relevant judgments \smash{$\hat{\mathcal{R}}_{L_{1:n}}$} for the top-$n$ items in the ranked list $L$, as shown in Figure~\ref{fig:precision}.
Note that in this case, $n=k$.
The following is an example to compute \ac{RR}@k:
\begin{equation}
RR@k=1/\min_{i}\{\hat{r}_i>0\}, 
\label{define:rr}
\end{equation}
where $0<i \leqslant k$.
For instance, as illustrated in Figure~\ref{fig:precision}, the first item in the ranked list is predicted as irrelevant, while the second item is predicted as relevant. In this case, the predicted $RR@k$ value would be 0.5.
$RR@k$ would be equal to 0 if there is no top-$k$ item that is predicted as relevant to the query $q$.
\begin{figure}[t]
     \centering
     \begin{tcolorbox}[notitle,boxrule=1pt,colback=gray!5,colframe=black, arc=2mm,width=\textwidth]
\sf\textbf{Instruction}: Please assess the relevance of the provided passage
to the following question. Please output ``Relevant'' or ``Irrelevant''.\\
    Question: \{question\} \\
    Passage: \{passage\} \\
    Output: Relevant/Irrelevant
    \end{tcolorbox}
     \caption{
    Prompt used by \acp{LLM} for automatic generation of relevance judgments.
     }
     \label{fig:prompt}
\end{figure}

\subsubsection{An approximation strategy to predict measures considering recall}
\label{sec:approx}

As the computation of a measure considering recall requires the information of all relevant items in the corpus $C$ for a given query $q$, we need to automatically assess every item in corpus $C$, which is infeasible due to the high computational cost.
To address this issue, we devise an approximation strategy for predicting an \ac{IR} measure considering recall, which only judges the top-$n$ ($n$ $\ll$ the total number of items in the corpus) items in the ranked list $L$ and uses the items predicted as relevant to approximate all relevant items in the corpus, to avoid the cost of judging the entire corpus.
%
\citet{frobe2023bootstrapped,moffat2017computing,lu2016effect} define \ac{nDCG}~\citep{jarvelin2002cumulated} at a cutoff $k$ as a recall-oriented \ac{IR} evaluation metric because it is normalized by a recall-oriented ``best possible'' ranking.\footnote{In this paper, we employ nDCG@10 and believe that nDCG@10 is a metric considering recall: Figure~\ref{fig:ndcg} illustrates that to reach saturation in predicting nDCG@10 values for ANCE and BM25, judgments up to the top 100 and 200 retrieved items are needed, respectively. If it were a precision-based metric, saturation could be achieved by judging around 10 items.}
nDCG@10 is also the most primary official \ac{IR} evaluation metric in TREC-DL 19--22~\citep{craswell2022,2021craswell,craswell2020,craswell2019}.
Thus, here we show an example of predicting nDCG@$k$~\citep{jarvelin2002cumulated}, formally: 
\begin{equation}
\begin{split}
nDCG@k={DCG@k}/{IDCG@k},
\end{split}
\label{ndcg}
\end{equation}
where ${DCG@k}$ can be computed easily using the generated relevance judgments for the top-$k$ items in the ranked list $L$, namely:\footnote{Note that we consider the definition of $DCG@k$ for binary relevance labels.}
\begin{equation}
\begin{split}
DCG@k=\hat{r}_1+\sum_{i=2}^{k} \hat{r}_i/\log_{2} i.
\end{split}
\label{dcg}
\end{equation}
$IDCG@k$ is the ideal ranked list with $k$ items, which requires knowing all the relevant items in the corpus $C$. 
We approximate all relevant items in the corpus by considering the items that are predicted as relevant at the top-$n$ ranks in the ranked list $L$, and compute $IDCG@k$ based on that.
First, we reorder the \ac{LLM}-generated relevant judgments \smash{$\hat{\mathcal{R}}_{L_{1:n}}=[\hat{r}_1, \dots, \hat{r}_i, \dots, \hat{r}_n]$} for the ranked list $L$ into \smash{$\hat{\mathcal{R}}_{iL_{1:n}}=[\hat{ir}_1, \dots, \hat{ir}_i, \dots, \hat{ir}_n]$} in descending order of predicted relevance;
then, we compute $IDCG@k$ based on \smash{$\hat{\mathcal{R}}_{iL_{1:n}}$}, namely: 
\begin{equation}
\begin{split}
IDCG@k=\hat{ir}_1+\sum_{i=2}^{k}\hat{ir}_i/\log _{2} i.
\end{split}
\label{idcg}
\end{equation}

\section{Experimental setup}
\label{sec:setup}
\subsection{Research questions}
In this section, we study the following research questions:
\begin{enumerate}[label=\textbf{RQ\arabic*}]
\item To what extent does \ourmodel improve \ac{QPP} quality for lexical and neural rankers in terms of RR@10, a precision-oriented \ac{IR} metric, compared to state-of-the-art baselines? \label{RQ1} \looseness=-1
\item To what extent does \ourmodel improve \ac{QPP} quality for lexical and neural rankers in terms of nDCG@10, an \ac{IR} metric that not only considers precision but also takes recall into account, compared to state-of-the-art baselines? \label{RQ2} 
\item How does judging depth in a ranked list affect the prediction of nDCG@10, an \ac{IR} metric that considers both precision and recall? In other words, how does varying the number of top-ranked documents submitted for relevance judgments impact \ac{QPP} quality? \label{RQ3}

%

\item To what extent do fine-tuning and the choice of \acp{LLM} affect the quality of generated relevance judgments and \ac{QPP}? \label{RQ4}




\end{enumerate}

\noindent%
\subsection{Datasets}
We experiment with 4 widely-used \ac{IR} datasets from the TREC 2019--2022 deep learning (TREC-DL)
tracks~\citep{craswell2022,2021craswell,craswell2020,craswell2019}.
These datasets provide relevance judgments in multi-graded relevance scales per query.
TREC-DL 19, 20, 21 and 22 have 43, 54, 53 and 76 queries, respectively.
TREC-DL 19/20 and TREC-DL 21/22 are based on the MS MARCO V1 and MS MARCO V2 passage ranking collections respectively. 
In the V1 edition, the corpus comprises 8.8 million passages while the V2 edition has over 138 million passages.

%

\subsection{Retrieval approaches}
We consider BM25~\cite{robertson1995okapi} as a lexical ranker; we also consider ANCE~\cite{xiong2021approximate} and TAS-B~\cite{hofstatter2021efficiently} as neural-based dense retrievers.
To increase the comparability and reproducibility of our paper, we get the retrieval results of both rankers using the publicly available resource from Pyserini~\citep{lin2021pyserini}.
We get BM25's retrieval result with top-1000 retrieved items per query on the TREC-DL 19--22 datasets using the default parameters ($k1=0.9$, $b=0.4$).
BM25's actual nDCG@10 values are 0.506, 0.480, 0.446 and 0.269 on TREC-DL 19, 20, 21 and 22, respectively.
We get the retrieval results of ANCE and TAS-B with top-1000 retrieved items per query on TREC-DL 19--20, using the publicly available dense vector index of ANCE on MS MARCO V1.
ANCE's actual nDCG@10 values are 0.645 and 0.646 on TREC-DL 19 and 20, respectively; TAS-B's actual nDCG@10 values are 0.721 and 0.685 on TREC-DL 19 and 20, respectively.
We rely on the publicly available dense vector index of ANCE/TAS-B; at the time of writing, there is no dense vector index of ANCE/TAS-B publicly available on MS MARCO V2 for TREC-DL 21 and 22.\footnote{Building the dense vector index on MS MARCO V2 with over 138 million passages is resource-intensive and beyond the scope of our work.}

\subsection{QPP baselines}
\label{baselines}
We consider three groups of baselines: unsupervised post-retrieval \ac{QPP} methods, supervised post-retrieval \ac{QPP} methods, and the \ac{LLM}-based \ac{QPP} methods.
Specifically, we consider the following unsupervised \ac{QPP} approaches that already showed high correlation with actual retrieval performance in previous work:
\begin{itemize}[leftmargin=*]
\item  Clarity~\cite{cronen2002predicting} computes the KL divergence between language models~\citep{lavrenko2001relevance} induced from the top-$k$ items in a ranked list and the corpus. 
\item  Weighted information gain (WIG)~\cite{zhou2007query} calculates the difference between retrieval scores of the top-$k$ items in a ranked list and the retrieval score of the entire corpus.
%
\item Normalized query commitment (NQC)~\cite{shtok2012predicting} calculates the standard deviation of retrieval scores of the top-$k$ items in a ranked list to a query; the standard deviation is normalized by the retrieval score of the entire corpus to the query.
\item $\sigma_{max}$~\citep{perez2010standard} computes the standard deviation of retrieval scores from the first item to each point in a ranked list and outputs the maximum standard deviation. 
%
\item n($\sigma_{x\%}$)~\citep{cummins2011improved} calculates the standard deviation for each query by considering the items whose retrieval scores are at least $x\%$ of the top retrieval score in a ranked list.
\item Score magnitude and variance (SMV)~\cite{tao2014query} considers both the magnitude of retrieval scores~(WIG) and their variance~(NQC).
\item UEF(NQC)~\citep{shtok2010using} uses a pseudo-effective reference list to improve \ac{QPP} quality; we follow \citep{arabzadeh2023noisy,datta2022pointwise,arabzadeh2021bert} to use NQC as a base predictor.
\item RLS(NQC)~\citep{roitman2017enhanced} generates and selects both pseudo-effective and pseudo-ineffective reference lists; we use NQC as a base predictor because \citet{roitman2017enhanced} show that RLS works better with NQC.
%
\item \ac{QPP-PRP}~\citep{singh2023unsupervised} measures the degree to which a pairwise neural re-ranker (DuoT5~\citep{spradeep2021expando}) agrees with the ranked list for the query. 
%

\item Dense-QPP~\citep{arabzadeh2023noisy} is robustness-based and designed for dense retrievers only: it injects noise neural representation of the given query, and then measures the similarity between ranked lists for the original query and perturbed query representations. 
Note that Dense-QPP~\citep{arabzadeh2023noisy} is designed for predicting the ranking quality of neural-based retrievers; it cannot predict the ranking quality of BM25.
\end{itemize}  

\noindent 
Since studies show that BERT-based post-retrieval supervised \ac{QPP} methods~\citep{hashemi2019performance,arabzadeh2021bert,datta2022pointwise,chen2022groupwise} perform better than their neural-based counterparts, we only consider BERT-based supervised \ac{QPP} approaches:
\begin{itemize}[leftmargin=*]
\item NQA-QPP~\citep{hashemi2019performance} is a regression-based method, which predicts a \ac{QPP} score by using BERT representations for the query and query-item pairs, and the standard deviation of retrieval scores.
\item BERTQPP~\citep{arabzadeh2021bert} is a regression-based method, which predicts a \ac{QPP} score by using BERT representations for the query and the top-ranked item. We use the cross-encoder version of BERTQPP because of its promising results.
\item qppBERT-PL~\citep{datta2022pointwise} first splits the ranked list into chunks, predicts the number of relevant items in each chunk, and calculates a weighted average of the number of relevant items in all chunks.
\item \ac{M-QPPF}~\citep{khodabakhsh2023learning} is also regression-based and models \ac{QPP} and document ranking jointly, by adopting a shared BERT layer to learn representations for query-document pairs, and using two layers to model \ac{QPP} and document ranking, respectively.
\end{itemize}  

\begin{figure}[t]
     \centering
     \begin{tcolorbox}[notitle,boxrule=1pt,colback=gray!5,colframe=black, arc=2mm,width=\textwidth]
\sf\textbf{Instruction}: Evaluate the relevance of the ranked list of passages to the given query by providing a numerical score between 0 and
1. A score of ``1'' indicates that the ranked passages are highly
relevant to the query, while a score of ``0'' means no relevance between the passages and the query.\\
Query: \{~\}\\
Passage 1: \{~\}\\
Passage 2: \{~\}\\
\ldots \\
Passage $k$: \{~\} \\
Output:
\end{tcolorbox}

     \caption{
    Prompt used by QPP-LLM.
     }
     \label{fig:prompt_baseline}
\end{figure}

\noindent%
While to the best of our knowledge there is no \ac{LLM}-based QPP method yet, to have a fair comparison with LLM-based approaches, we propose two \ac{LLM}-based \ac{QPP} baselines.
Research on using \acp{LLM} for arithmetic tasks shows that LLaMA treats numbers as distinct tokens and can understand and generate numerical values~\citep{liu2023goat}.
Inspired by this, we prompt LLaMA-7B to directly generate a numerical score given a query and the ranked list with $k$ passages for the query; the prompt is shown in Figure~\ref{fig:prompt_baseline}. 
We consider two variants:
\begin{itemize}[leftmargin=*]
\item QPP-LLM (few-shot) uses \acf{ICL} and inserts several demonstration examples after the instruction in the prompt; each example is composed of a query, $k$ passages and the actual performance in terms of an \ac{IR} evaluation measure. 
\item QPP-LLM (fine-tuned) fine-tune LLaMA-7B to learn to directly generate numerical values of an \ac{IR} metric, similar to the way other regression-based supervised \ac{QPP} methods are trained. 
\end{itemize}



\subsection{QPP evaluation and target IR evaluation measures}
We follow established best practices~\citep{datta2022pointwise,zamani2018neural,carmel2010estimating,hauff2008survey,cronen2002predicting} to evaluate \ac{QPP} by measuring linear correlation by Pearson's $\rho$ as well as ranked-based correlation through Kendall's $\tau$ correlation coefficients between the actual and predicted performance of a query set.
%
As for target \ac{IR} metrics, we consider the two primary official \ac{IR} metrics used in TREC DL 19--22~\citep{craswell2022,2021craswell,craswell2020,craswell2019}, RR@10 (precision-oriented) and nDCG@10 (considering recall); recent \ac{QPP} studies~\citep{arabzadeh2023noisy,faggioli2023towards,khodabakhsh2023learning} consider either or both of these metrics as their target metrics.
Following~\citep{datta2022pointwise}, we use relevance scale $\geq$ 2 as positive to compute actual binary \ac{IR} measures (e.g., RR).
When calculating correlation for nDCG@10, the actual values of nDCG@10 are calculated by human-labeled and multi-graded relevance judgments, while the nDCG@10 values predicted by \ourmodel are based on its generated binary judgments.


\subsection{Implementation details}
\label{implementation}
For all unsupervised \ac{QPP} baselines, we tune the hyper-parameters for predicting the ranking quality of a ranker (either BM25 or ANCE) on TREC-DL 19 (TREC-DL 21) based on Pearson's~$\rho$ correlation for predicting the ranking quality of the same ranker on TREC-DL 20 (TREC-DL 22), and vice versa.
We select the cut-off value $k$ for Clarity, NQC, WIG, SMV and so on from $\{5,10,15,20,25,50,100,300,500,1000\}$.
n($\sigma_{x\%}$) has a hyper-parameter $x$, which we choose from the set $\{0.25, 0.4, 0.5, 0.6, 0.75, 0.9\}$.

To predict the performance of a certain ranker (any of BM25, ANCE, or TAS-B), we train all supervised \ac{QPP} baselines based on the ranked list returned by the target ranker.
To predict a certain \ac{IR} evaluation measure, regression-based methods~\citep{khodabakhsh2023learning,arabzadeh2021bert,hashemi2019performance} are trained to learn to output the target evaluation measure during training.
However, our preliminary result shows that training supervised \ac{QPP} baselines, especially for regression-based supervised methods~\citep{khodabakhsh2023learning,arabzadeh2021bert,hashemi2019performance}, on the training set of MS MARCO V1 leads to inferior \ac{QPP} quality for predicting the performance of the neural rankers (ANCE and TAS-B).
We hypothesize that this is because they were originally trained on the training set of MS MARCO V1~\citep{xiong2021approximate,hofstatter2021efficiently}, and so the ranked list returned by them on the training set of MS MARCO V1 would have higher quality than the ranked list returned by them on the evaluation sets; therefore, supervised \ac{QPP} methods that share the same training set as the neural rankers, tend to predict inflated performance on the evaluation sets, leading to degraded \ac{QPP} quality.
To solve the issue and ensure the consistency of the paper, we train all supervised \ac{QPP} methods (including \ourmodel) on the development set of MS MARCO V1 (6980 queries) for predicting the performance of BM25, ANCE or TAS-B.
We train all supervised \ac{QPP} methods for 5 epochs and pick the best checkpoint for predicting the performance of a ranker on TREC-DL 19 (TREC-DL 21) based on Pearson's $\rho$ correlation for predicting the performance of the same ranker on TREC-DL 20 (TREC-DL 22) and vice versa.
All supervised \ac{QPP} baselines use bert-base-uncased,\footnote{\url{https://github.com/huggingface/transformers}} 
a constant learning rate (0.00002), and the Adam optimizer~\citep{kingma2014adam}.

For QPP-LLM, we prompt LLaMA-7B with the top-$k$  retrieved items, where $k$ is set to 10.
For QPP-LLM (few-shot), we randomly sample demonstration examples from the development set of MS MARCO V1; our preliminary experiments show that sampling 2 demonstrations works best.
For QPP-LLM (fine-tuned), we fine-tune LLaMA-7B using \ac{PEFT} as \ourmodel fine-tunes \acp{LLM}.

We equip \ourmodel with an \ac{LLM} for judging relevance.
We use a recent \ac{PEFT} method, 4-bit QLoRA~\citep{dettmers2023qlora}, to fine-tune an \ac{LLM}.
To maintain a comparable setup with the baselines, we fine-tune an \ac{LLM} for 5 epochs on the development set of MS MARCO V1.
Note that we use LLaMA-7B for BM25 and ANCE, and Mistral-7B-Instruct-v0.3 for TAS-B.
The training of judging relevance needs positive and negative items per query.
For positive items, we use the items annotated as relevant in \textit{qrels} per query; we randomly sample one negative item from the ranked list~(1,000 items) returned by BM25 per query.
%
There are 6,980 queries in our training set.
Each each query may have multiple relevant items annotated in the \textit{qrels}, and has one negative item we sampled.
As a result, we have 7,437 positive training examples and 6,980 negative training examples.
All experiments are conducted on an NVIDIA A100 GPU (40GB).

One might argue why we choose QLoRA fine-tuning instead of distilling an oracle model into a smaller model.~\citep{sun2023chatgpt}. 
The decision is based on the following three reasons.
First, model distillation requires an existing oracle model, such as GPT-4, for relevance prediction. However, this work focuses on exclusively using open-source LLMs, avoiding using powerful commercial models like GPT-4 to ensure reproducibility and deterministic outputs.
Second, one might wonder why we did not distill larger open-source LLMs into smaller models. 
As demonstrated in Figure~\ref{fig:llm}, a 1-billion-parameter Llama model fine-tuned using QLoRA on human-labeled relevance judgments outperforms a 70-billion-parameter Llama model using few-shot prompting. 
Therefore, if we choose to distill the 70-billion-parameter model into a smaller model, the performance of the distilled model would be inferior to that achieved by the 1-billion-parameter model fine-tuned via QLoRA, because the performance of distilled model is inherently limited by the larger model's capabilities.
Third, we have a large amount of human-labeled relevance judgments available. 
Directly using these labels to fine-tune LLMs via QLoRA allows us to make the most efficient use of this data.

\begin{table*}[!t]
\caption{
Correlation coefficients (Pearson's $\rho$ and Kendall's $\tau$) between actual retrieval quality, in terms of RR@10, of BM25 and performance predicted by \ourmodel/baselines, on TREC-DL 19--22.
$^*$ indicates statistically significant correlation coefficients ($p$-value $< 0.05$). 
$^\dagger$ indicates the statistically significant improvement of \ourmodel compared to all the baselines (paired $t$-test; $p$-value $< 0.001$ with Bonferroni correction for multiple testing).
The best value in each column is marked in \textbf{bold}.
$n$ denotes \ourmodel's judgment depth in a ranked list.
}
\label{tab:precision:lexical}
\setlength{\tabcolsep}{1mm}
\begin{tabular}{l ll ll ll ll}
\toprule
& \multicolumn{8}{c}{Ranker: BM25} \\
 \cmidrule(lr){2-9}
& \multicolumn{2}{c}{TREC-DL 19} & \multicolumn{2}{c}{TREC-DL 20} & \multicolumn{2}{c}{TREC-DL 21} & \multicolumn{2}{c}{TREC-DL 22} \\
 \cmidrule(lr){2-3}  \cmidrule(lr){4-5}  \cmidrule(lr){6-7}   \cmidrule(lr){8-9}
 QPP method
  & \multicolumn{1}{c}{P-$\rho$} & \multicolumn{1}{c}{K-$\tau$}  & \multicolumn{1}{c}{P-$\rho$} & \multicolumn{1}{c}{K-$\tau$}   & \multicolumn{1}{c}{P-$\rho$} & \multicolumn{1}{c}{K-$\tau$}  & \multicolumn{1}{c}{P-$\rho$} & \multicolumn{1}{c}{K-$\tau$} \\
\midrule    
 Clarity & {0.135} & {0.028}  & {0.050} & {0.021}   & {0.183} & {0.161}    & 0.253$^*$ &  {0.099}   \\ 
 WIG & {0.113} & {0.164}   & 0.286$^*$ & 0.218$^*$ & {0.237} & 0.206$^*$  & {0.029} & {0.082}   \\  
 NQC & {0.194} & {0.117}  & {0.152}  & {0.191}   & {0.227} & {0.195}  & {0.223} & {0.048}  \\
 $\sigma_{max}$ & {0.195} & {0.164}  & {0.200}  & 0.211$^*$   & 0.278$^*$  & {0.174}   & {0.038}  & {0.048}  \\  
 n($\sigma_{x\%}$)  & {0.144}  & {0.181}   & {0.187}  & {0.123}    & {0.127} & {0.140}    & {0.169} & {0.113}  \\
 SMV  & {0.141}  & {0.097}   & {0.126} & {0.193}   & {0.240} & {0.189}  & 0.227$^*$ & {0.094}   \\ 
UEF(NQC)  & 0.235 & 0.256$^*$ & 0.270$^*$ & 0.211$^*$ & 0.231 & 0.111 & 0.216  & 0.065  \\
RLS(NQC)   & 0.272  & 0.122  & 	0.290$^*$ & 0.193  & 0.234 & 0.195  & 	0.224 & 0.095   \\
\ac{QPP-PRP} & {0.292} & {0.189} & {0.163} & {0.184} & \llap{-}0.080 & \llap{-}0.017 &  {0.122} &  {0.091}    \\
\midrule       
 NQA-QPP & {0.181} & {0.122} & {0.062} & {0.069}  & {0.161} & {0.163}   & {0.224} & 0.177$^*$   \\   
 BERTQPP & {0.281} & {0.136}   & {0.237}  & {0.155}  &  {0.206}  & {0.134}   & {0.148}  & {0.122}   \\  
  qppBERT-PL & {0.145}  & {0.138}  & {0.166} & {0.152}  & 0.339$^*$ & 0.244$^*$   & {0.131}  & 0.206$^*$     \\
 \ac{M-QPPF} & 0.317$^*$  & 0.208  & 0.335$^*$  & 0.273$^*$ & 0.282$^*$  & 0.209$^*$ & 0.161 & 0.187$^*$  \\ 
 \midrule  
QPP-LLM (few-shot) & {0.008}  & {0.003} & \llap{-}0.081  & \llap{-}0.129 & \llap{-}0.053 & \llap{-}0.053 & \llap{-}0.241 & \llap{-}0.155 \\
QPP-LLM (fine-tuned) & {0.171} & {0.158} & {0.228} & {0.206} & {0.030} & {0.099} & \llap{-}0.038 & {0.009}  \\
\midrule 
\ourmodel ($n=10$) & \textbf{0.538}$^\dagger$$^*$ & \textbf{0.486}$^\dagger$$^*$  & \textbf{0.560}$^\dagger$$^*$ & \textbf{0.475}$^\dagger$$^*$  & \textbf{0.524}$^\dagger$$^*$ & \textbf{0.435}$^\dagger$$^*$   &\textbf{0.350}$^\dagger$$^*$ & \textbf{0.262}$^\dagger$$^*$  \\

  \bottomrule
\end{tabular}
\end{table*}

\begin{table*}[!th]
\caption{
Correlation coefficients (Pearson's $\rho$ and Kendall's $\tau$) between actual retrieval quality, in terms of RR@10, of ANCE/TAS-B and performance predicted by \ourmodel/baselines, on TREC-DL 19 and 20.
$^*$ indicates statistically significant correlation coefficients ($p$-value $< 0.05$). 
$^\dagger$ indicates the statistically significant improvement of \ourmodel compared to all the baselines (paired $t$-test; $p$-value $< 0.001$ with Bonferroni correction for multiple testing).
The best value in each column is marked in \textbf{bold}.
$n$ denotes \ourmodel's judgment depth in a ranked list.
}
\label{tab:precision:neural}
\setlength{\tabcolsep}{1mm}
\begin{tabular}{l ll ll ll ll} 
\toprule
& \multicolumn{4}{c}{Ranker: ANCE} & \multicolumn{4}{c}{Ranker: TAS-B}\\
 \cmidrule(lr){2-5}   \cmidrule(lr){6-9} 
& \multicolumn{2}{c}{TREC-DL 19} & \multicolumn{2}{c}{TREC-DL 20}  & \multicolumn{2}{c}{TREC-DL 19} & \multicolumn{2}{c}{TREC-DL 20} \\
 \cmidrule(lr){2-3}  \cmidrule(lr){4-5}  \cmidrule(lr){6-7} \cmidrule(lr){8-9} 
 QPP method
  & \multicolumn{1}{c}{P-$\rho$} & \multicolumn{1}{c}{K-$\tau$}  & \multicolumn{1}{c}{P-$\rho$} & \multicolumn{1}{c}{K-$\tau$}    & \multicolumn{1}{c}{P-$\rho$} & \multicolumn{1}{c}{K-$\tau$}  & \multicolumn{1}{c}{P-$\rho$} & \multicolumn{1}{c}{K-$\tau$}  \\
\midrule   
 Clarity  & \llap{-}0.078  & \llap{-}0.012  & \llap{-}0.074  & \llap{-}0.048  &  \llap{-}0.212 & \llap{-}0.148  & 0.148 & 0.133 \\ 
 WIG  & 0.313$^*$  & {0.228}  & {0.059}  & {0.048} & \llap{-}0.066 & \llap{-}0.125 & 0.024 & 0.020 \\  
 NQC & 0.350$^*$  & {0.200}  & {0.145} & {0.112} & 0.248 & 0.213  & 0.260 & 0.194 \\ 
 $\sigma_{max}$  & 0.384$^*$ & 0.287$^*$ & {0.171}  & {0.118} & 0.015 & 0.021 & 0.312$^*$ & 0.245$^*$ \\    
 n($\sigma_{x\%}$)    & {0.200}  & {0.176} & \llap{-}0.008 & {0.022} &  \llap{-}0.030 & \llap{-}0.079   &  0.080 & 0.086 \\ 
 SMV   & 0.352$^*$ & 0.256$^*$ & {0.182} & {0.161}  & 0.249 & 0.205 & 0.263 & 0.198 \\  
UEF(NQC)  & 0.340$^*$  & 0.260$^*$  & 0.131  & 0.108 & 0.260 & 0.228 & 0.281$^*$  & 0.213 \\ 
RLS(NQC)   & 0.359$^*$  & 0.273$^*$  & 	0.178  & 0.139 & 0.257 & 0.217  & 0.283$^*$  & 0.217$^*$  \\ 
\ac{QPP-PRP}  & {0.259}  & 0.246 & {0.100} & \llap{-}0.008  & 0.155 & 0.113  &  0.203 & 0.116 \\  
 Dense-QPP   & 0.452$^*$  & 0.280$^*$  &  0.209 & 0.139 & 0.251 & 0.213  & 0.146  & 0.012 \\  
\midrule       
 NQA-QPP   & \llap{-}{0.026}  & \llap{-}0.009  & \llap{-}0.059 & \llap{-}0.080  & 0.172 & 0.144 & \llap{-}0.058 & \llap{-}0.075 \\ 
 BERTQPP  & 0.330$^*$ & {0.214}  & {0.046} & \llap{-}0.012 & 0.202 & 0.194 & 0.077 & 0.037 \\  
 qppBERT-PL   & {0.092} & {0.025}  & \llap{-}0.224 & \llap{-}0.218    & 0.276 & 0.269 & 0.004 &  \llap{-}0.002 \\ 
 \ac{M-QPPF}  & 0.292  & 0.200 & 0.068  & 0.038 & 0.277 & 0.236  & 0.103  & 0.022 \\ 
\midrule  
QPP-LLM (few-shot) & \llap{-}0.008  & {0.005}  & \llap{-}0.226 & \llap{-}0.207 &  \llap{-}0.080 & 0.002   &  0.054 & \llap{-}0.024 \\ 
QPP-LLM (fine-tuned)  & \llap{-}0.073  & {0.011} & \llap{-}0.022 & {0.069} & 0.155 & 0.113 & 0.043  & \llap{-}0.020 \\ 
\midrule 
\ourmodel ($n=10$)  & \textbf{0.567}$^\dagger$$^*$ & \textbf{0.440}$^\dagger$$^*$  & \textbf{0.293}$^\dagger$$^*$ & \textbf{0.257}$^\dagger$$^*$ & \textbf{0.538}$^\dagger$$^*$ & \textbf{0.481}$^\dagger$$^*$  & \textbf{0.356}$^\dagger$$^*$ & \textbf{0.289}$^\dagger$$^*$ \\
\bottomrule
\end{tabular}
\end{table*}

\section{Results}
\subsection{Predicting a precision-oriented IR measure}
\label{subsec:precision}

To answer \ref{RQ1}, we compare \ourmodel and all baselines in predicting the performance of BM25, ANCE and TAS-B w.r.t.\ a widely-used precision-oriented metric, RR@10; see Tables~\ref{tab:precision:lexical} and \ref{tab:precision:neural}.
%
We have three main observations.

First, our proposed method, \ourmodel, outperforms all baselines in terms of both correlation coefficients on all datasets when predicting the performance of all rankers. 
In particular, we observe that \ourmodel outperforms \ac{QPP-PRP}~\citep{singh2023unsupervised}, which is a recently proposed baseline by 84\% (0.292 vs.~0.538) in terms of Pearson's $\rho$ when predicting RR@10 for BM25 on TREC-DL 19.

Second, QPP-LLM (few-shot) gets the worst result compared to other approaches. 
While QPP-LLM (fine-tuning) performs slightly better than QPP-LLM (few-shot), its performance is still limited in most cases. 
This indicates that it is ineffective for an \ac{LLM} to model \ac{QPP} in a straightforward way of directly predicting a score.

Third, there is no clear winner among the baselines, and the performance of baselines shows a bigger variance than \ourmodel across different datasets and rankers.
E.g., the unsupervised method WIG achieves a good result among baselines for assessing BM25 on TREC-DL 20, while it gets nearly zero correlation coefficients on TREC-DL 22 when assessing BM25.
Conversely, \ourmodel consistently achieves the best performance across datasets and rankers, thus showing robust performance.


\begin{table*}[t]
\caption{
Correlation coefficients (Pearson's $\rho$ and Kendall's $\tau$) between actual retrieval quality, in terms of nDCG@10, of BM25 and performance predicted by \ourmodel/baselines, on TREC-DL 19--22.
$n$ denotes \ourmodel's judgment depth in a ranked list.
$^*$ indicates statistically significant correlation coefficients ($p$-value $< 0.05$). 
$^\dagger$ indicates the statistically significant improvement of \ourmodel ($n$=200) compared to all the baselines (paired $t$-test; $p$-value $< 0.001$ with Bonferroni correction for multiple testing).
The best value in each column is marked in \textbf{bold}.
}
\label{tab:recall:lexcial}
\setlength{\tabcolsep}{1mm}
\begin{tabular}{l ll ll ll ll}
\toprule
& \multicolumn{8}{c}{Ranker: BM25} \\
 \cmidrule(lr){2-9}
& \multicolumn{2}{c}{TREC-DL 19} & \multicolumn{2}{c}{TREC-DL 20} & \multicolumn{2}{c}{TREC-DL 21} & \multicolumn{2}{c}{TREC-DL 22}  \\
 \cmidrule(lr){2-3}  \cmidrule(lr){4-5}  \cmidrule(lr){6-7}   \cmidrule(lr){8-9}
 QPP method
  & \multicolumn{1}{c}{P-$\rho$} & \multicolumn{1}{c}{K-$\tau$}  & \multicolumn{1}{c}{P-$\rho$} & \multicolumn{1}{c}{K-$\tau$}   & \multicolumn{1}{c}{P-$\rho$} & \multicolumn{1}{c}{K-$\tau$}  & \multicolumn{1}{c}{P-$\rho$} & \multicolumn{1}{c}{K-$\tau$}  \\
\midrule    
 Clarity & {0.091} & {0.056}    & 0.358$^*$ & 0.250$^*$  & {0.137} & {0.078}   & {0.202}  & {0.090}   \\
 WIG & 0.520$^*$ & 0.331$^*$ &  0.615$^*$  & 0.423$^*$ & 0.311$^*$ & 0.281$^*$   & 0.350$^*$  & 0.249$^*$  \\  
 NQC & 0.468$^*$ & 0.300$^*$    & 0.508$^*$ & 0.401$^*$  & {0.134} & 0.221$^*$   & 0.360$^*$  & 0.156$^*$ \\ 
 $\sigma_{max}$ & 0.478$^*$  & 0.327$^*$ & 0.529$^*$ & 0.440$^*$   & 0.298$^*$ & 0.258$^*$    & 0.142$^*$  & 0.196$^*$   \\  
 n($\sigma_{x\%}$)  & 0.532$^*$ & 0.311$^*$   & 0.622$^*$ & 0.443$^*$   & 0.328$^*$ & 0.234$^*$    & 0.336$^*$ & 0.228$^*$  \\
 SMV   & 0.376$^*$ & 0.271$^*$  & 0.463$^*$ & 0.383$^*$  & 0.327$^*$  & 0.236$^*$  & 0.338$^*$ & 0.155$^*$  \\  
UEF(NQC)  & 0.499$^*$ & 0.322$^*$ & 0.517$^*$ & 0.356$^*$ & 0.153 & 0.232$^*$ & 0.311$^*$ & 0.145  \\
RLS(NQC)   & 0.469$^*$  & 0.169 & 0.522$^*$  & 0.376$^*$ & 0.272$^*$ & 0.223$^*$ & 0.337$^*$ & 0.157$^*$ \\
\ac{QPP-PRP} & 0.321  & {0.181} & {0.189} & {0.157} & {0.027} & {0.004}  &  {0.077} &  {0.012}   \\
\midrule 
 NQA-QPP &  {0.210}  & {0.147}  &  {0.244} & 0.210$^*$  & 0.286$^*$ & 0.201$^*$   & 0.312$^*$ & 0.194$^*$    \\ 
 BERTQPP  & 0.458$^*$ & {0.207}  & 0.426$^*$ & 0.300$^*$  &  0.351$^*$ & 0.223$^*$   & 0.369$^*$ & 0.229$^*$    \\
 qppBERT-PL & {0.171}  & {0.175}   & 0.410$^*$ & 0.279$^*$ & 0.277$^*$ & {0.182}   & 0.300$^*$ & 0.242$^*$ \\
 \ac{M-QPPF} & 0.404$^*$ & 0.254$^*$ & 0.435$^*$ & 0.297$^*$  & 0.265  & 0.226$^*$ & 0.345$^*$ & 0.204$^*$  \\ 
 \midrule   
QPP-LLM (few-shot) &  \llap{-}0.024 & \llap{-}0.031 & {0.167} & {0.138} & {0.238}  & {0.201}  & \llap{-}0.073 & \llap{-}0.077   \\
QPP-LLM (fine-tuned) & 0.313$^*$ & {0.215}  & 0.309$^*$ & 0.254$^*$ & {0.264}  & {0.198}  & \llap{-}0.075  & \llap{-}0.009 \\
\midrule 

\ourmodel ($n=200$)  & \textbf{0.724}$^\dagger$$^*$  &  0.474$^\dagger$$^*$  & \textbf{0.638}$^\dagger$$^*$ & \textbf{0.469}$^\dagger$$^*$     &  0.546$^\dagger$$^*$ & 0.435$^\dagger$$^*$   & \textbf{0.388}$^*$ & \textbf{0.251}$^*$   \\
\midrule 
\ourmodel ($n=10$) & 0.605$^*$ & \textbf{0.482}$^*$ & 0.490$^*$ & 0.323$^*$  & 0.462$^*$  & 0.350$^*$ & 0.316$^*$ & 0.245$^*$ \\
\ourmodel ($n=100$) & 0.712$^*$ & 0.472$^*$ & 0.609$^*$ & 0.457$^*$ & 0.545$^*$ & 0.427$^*$ & 0.332$^*$  & 0.246$^*$ \\
\ourmodel ($n=1,000$) & 0.715$^*$ & 0.477$^*$ & 0.627$^*$ & 0.459$^*$ & \textbf{0.547}$^*$ & \textbf{0.436}$^*$ & \textbf{0.388}$^*$  & \textbf{0.251}$^*$  \\
  \bottomrule
\end{tabular}
\end{table*}

\begin{table*}[t]
\caption{
Correlation coefficients (Pearson's $\rho$ and Kendall's $\tau$) between actual retrieval quality, in terms of nDCG@10, of ANCE/TAS-B and performance predicted by \ourmodel/baselines, on TREC-DL 19 and 20.
$n$ denotes \ourmodel's judgment depth in a ranked list.
$^*$ indicates statistically significant correlation coefficients ($p$-value $< 0.05$). 
$^\dagger$ indicates the statistically significant improvement of \ourmodel ($n$=200) compared to all the baselines (paired $t$-test; $p$-value $< 0.001$ with Bonferroni correction for multiple testing).
The best value in each column is marked in \textbf{bold}.
%
%
}
\label{tab:recall:neural}
\setlength{\tabcolsep}{1mm}
\begin{tabular}{l ll ll ll ll}
\toprule  
& \multicolumn{4}{c}{Ranker: ANCE} & \multicolumn{4}{c}{Ranker: TAS-B}\\
 \cmidrule(lr){2-5}  \cmidrule(lr){6-9}
& \multicolumn{2}{c}{TREC-DL 19} & \multicolumn{2}{c}{TREC-DL 20} & \multicolumn{2}{c}{TREC-DL 19} & \multicolumn{2}{c}{TREC-DL 20} \\
 \cmidrule(lr){2-3}  \cmidrule(lr){4-5} \cmidrule(lr){6-7} \cmidrule(lr){8-9}
 QPP method
  & \multicolumn{1}{c}{P-$\rho$} & \multicolumn{1}{c}{K-$\tau$}  & \multicolumn{1}{c}{P-$\rho$} & \multicolumn{1}{c}{K-$\tau$}   & \multicolumn{1}{c}{P-$\rho$} & \multicolumn{1}{c}{K-$\tau$}  & \multicolumn{1}{c}{P-$\rho$} & \multicolumn{1}{c}{K-$\tau$}    \\
\midrule    
 Clarity & \llap{-}0.088  & \llap{-}0.062  & \llap{-}0.091  & \llap{-}0.045  &  {0.153}  &  {0.049}  &  {0.162}  &   {0.087} \\ 
 WIG  & 0.515$^*$ & 0.368$^*$  & {0.218}  & {0.150}  &  {0.228} &   {0.146} &  {0.227}  &  {0.169} \\   
 NQC & 0.548$^*$ & 0.372$^*$  & 0.411$^*$ & 0.290$^*$  &  {0.330$^*$} &   {0.233$^*$} &  {0.406$^*$}  &  {0.264$^*$}  \\ 
 $\sigma_{max}$  & 0.455$^*$ & 0.339$^*$  & 0.403$^*$ & 0.288$^*$  &  {0.220} &  {0.126}  &  {0.428$^*$} &  {0.284$^*$} \\ 
 n($\sigma_{x\%}$)  & 0.388$^*$ & 0.315$^*$  & {0.103}  & {0.075} &  {\llap{-}0.008}  &  {\llap{-}0.031} &  {0.002} &   {\llap{-}0.020} \\ 
 SMV   & 0.496$^*$  & 0.359 & 0.380$^*$ & 0.283$^*$ &  {0.349$^*$} &  {0.253$^*$}  &  {0.425$^*$} &  {0.285$^*$} \\   
UEF(NQC)   & 0.548$^*$ & 0.372$^*$ & 0.413$^*$  & 0.290$^*$ &  {0.321$^*$} &  {0.246$^*$} &  {0.425$^*$}  &  {0.271$^*$} \\ 
RLS(NQC)   & 0.466$^*$ & 0.346$^*$ & 0.333$^*$ & 0.271$^*$ &  {0.314$^*$} &  {0.246$^*$} &  {0.404$^*$} &  {0.272$^*$} \\ 
\ac{QPP-PRP}  &  {0.129}  & {0.049} & {0.216} & {0.121} 
&  {0.220}  &  {0.126}  &  {0.267} &  {0.237$^*$} \\
Dense-QPP   & 0.565$^*$ & 0.389$^*$ & 0.419$^*$  & 0.318$^*$  &  {0.429$^*$}  &  {0.244$^*$ }&  {0.126} &  {0.012}  \\  
\midrule 
 NQA-QPP  & {0.089} & \llap{-}0.038 & {0.186} &  {0.113}    &   {\llap{-}0.020} &  {0.060} &  {0.031} &  {0.024} \\  
 BERTQPP  & {0.222} & {0.117}  & {0.137} & {0.089}   &  {0.043}  &  {0.027}  &  {0.178} &   {0.086} \\ 
 qppBERT-PL  & {0.116} & {0.098}  & \llap{-}0.119  & \llap{-}0.046  &  {0.304$^*$} &  {0.187} &  {0.057}  &  {0.057} \\ 
 \ac{M-QPPF} & 0.287  & 0.160  & 0.225  &  0.177 &  {0.163} &  {0.051}  &  {0.304$^*$} &  {0.171} \\ 
 \midrule   
QPP-LLM (few-shot) & {0.136} & {0.120} & \llap{-}0.130 & \llap{-}0.094  &  {\llap{-}0.020} &  {0.060}  &  {0.108} &  {0.048} \\ 
QPP-LLM (fine-tuned) & {0.203} & {0.117}  & {0.081} & {0.097} &  {0.262}   &   {0.195}  &  {0.162} &  {0.111} \\ 
\midrule 

\ourmodel ($n=200$)    & 0.712$^\dagger$$^*$  & 0.483$^\dagger$$^*$  & \textbf{0.457}$^\dagger$$^*$  & 0.343$^\dagger$$^*$ &   {0.501$^\dagger$$^*$} &  {0.346$^\dagger$$^*$} &  {0.449$^*$}  &  {0.315$^*$} \\  
\midrule 
\ourmodel ($n=10$)  & 0.624$^*$ & 0.406$^*$ & 0.306$^*$ & 0.238$^*$ &  {0.490$^*$} &  {0.309$^*$} &  {0.421$^*$} &  {0.290$^*$} \\ 
\ourmodel ($n=100$)  & \textbf{0.719}$^*$ & 0.489$^*$ & 0.456$^*$ & \textbf{0.355}$^*$ &  {0.501$^*$}  &  {0.336$^*$} &  {\textbf{0.450}$^*$} &  {\textbf{0.317}$^*$} \\
\ourmodel ($n=1,000$) & \textbf{0.719}$^*$  & \textbf{0.492}$^*$  & 0.447$^*$  & 0.321$^*$ &   {\textbf{0.505}$^*$} &  {\textbf{0.348}$^*$} &  {0.449$^*$} &  {0.315$^*$} \\ 
\bottomrule
\end{tabular}
\end{table*}

\subsection{Predicting an IR measure considering recall}
\label{subsec:recall}
%
To answer \ref{RQ2}, Tables~\ref{tab:recall:lexcial} and \ref{tab:recall:neural} list the performance of \ourmodel along with all the baselines on assessing BM25, ANCE and TAS-B in terms of nDCG@10.
For \ourmodel, we universally set the judging depth $n$ to 200 for all evaluation sets. 
The result reveals that by judging only 200 items per query, we can achieve state-of-the-art \ac{QPP} quality in terms of nDCG@10 for all rankers on all evaluation sets;
we will investigate the impact of judging depth on \ourmodel's performance in the next section.
Also, QPP-LLM (few-shot) and QPP-LLM (fine-tuning) are among the worst-performing baselines, showing that the \acp{LLM} struggle to generate numerical scores. 
Different from the results for \ref{RQ1}, most \ac{QPP} methods tend to perform better when predicting nDCG@10 than RR@10; this observation indicates that predicting RR@10 is a more challenging task.

\section{Analysis}

\begin{figure*}[t]
    \centering
    \begin{subfigure}{0.495\columnwidth}
        \includegraphics[width=\linewidth]{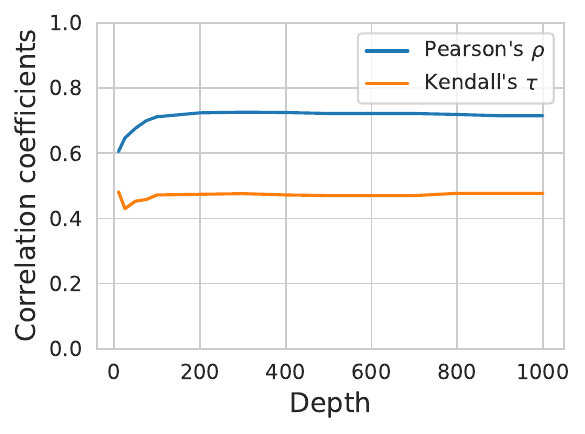}
        \caption{BM25 on TREC-DL 19}
        \label{fig:ndcg-dl19-bm25}
    \end{subfigure}
    \begin{subfigure}{0.495\columnwidth}
        \includegraphics[width=\linewidth]{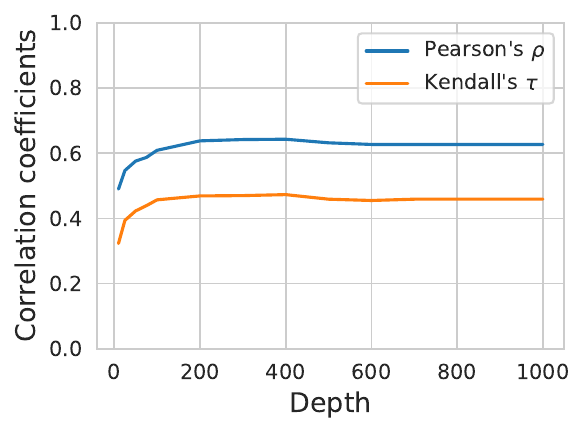}
        \caption{BM25 on TREC-DL 20}
        \label{fig:ndcg-dl20-bm25}
    \end{subfigure}
    \begin{subfigure}{0.495\columnwidth}
        \includegraphics[width=\linewidth]{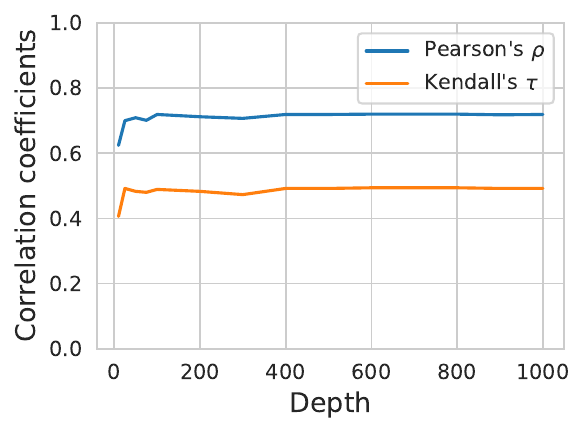}
        \caption{ANCE on TREC-DL 19}
        \label{fig:ndcg-dl19-ance}
    \end{subfigure}
    \begin{subfigure}{0.495\columnwidth}
        \includegraphics[width=\linewidth]{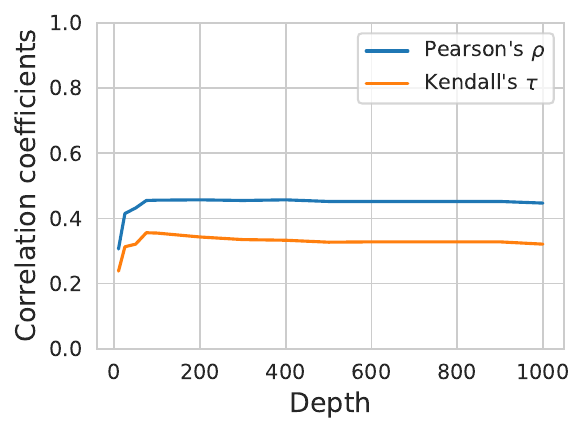}
        \caption{ANCE on TREC-DL 20}
        \label{fig:ndcg-dl20-ance}
    \end{subfigure}
    \caption{
    Relationship between the \ac{QPP} effectiveness of predicting nDCG@10 and the judging depth for a ranked list.
    }
    \label{fig:ndcg}
\end{figure*}

\subsection{Judging depth analysis} 
\label{sec:depth}

\ref{RQ3} examines how varying the number of top-ranked documents submitted for relevance judgments impacts \ac{QPP} quality.
To answer \ref{RQ3}, as detailed in Section~\ref{sec:approx}, for predicting IDCG, we devise an approximation strategy and use the items in the top $n$ ranks of the ranked list $L$ that are predicted as relevant by \ourmodel to approximate all the relevant items for a query in the corpus. 
To investigate the impact of the value of $n$ on the quality of the prediction, we investigate the relationship between the \ac{QPP} quality of predicting nDCG@10 and the judgment depth to answer the following question: \textit{What depth of relevance judgment $n$ do we need to consider to get a satisfactory performance for predicting nDCG@10?}
In Figure~\ref{fig:ndcg}, we plot the correlation coefficients between actual nDCG@10 values and nDCG@10 values predicted by \ourmodel for different judging depths in \{10, 25, 50, 75, 100, 200, 300, 400, 500, 600, 700, 800, 900, 1,000\} on TREC-DL 19 and 20.
We also show exact \ac{QPP} results with depths at 10, 100 and 1000 in Tables~\ref{tab:recall:lexcial} and \ref{tab:recall:neural}.

Tables~\ref{tab:recall:lexcial} and \ref{tab:recall:neural} reveal that, by judging only 10 items in the ranked list, we can already outperform all the baselines and achieve state-of-the-art \ac{QPP} quality on half of the evaluation sets we used, e.g., assessing BM25 on TREC-DL 19/21, ANCE on TREC-DL 19, and TAS-B on TREC-DL 19.
While judging deeper in the ranked list is essential for predicting recall-oriented measures, satisfactory \ac{QPP} quality is still attainable with a relatively shallow depth.
Moreover, Figure~\ref{fig:ndcg} illustrates that judging the top 200 items in a ranked list already reaches the saturation point for assessing BM25, i.e., there is no significant improvement by judging a higher number of items, while judging less than 100 top items reaches the saturation point for ANCE. 
We speculate that this is because ANCE has better retrieval quality than BM25, and more relevant items would appear earlier in the ranked list of ANCE than BM25; therefore, a shallower judging depth suffices to approximate all relevant items in the corpus. 
This emphasizes the need to consider retrieval quality when determining the optimal judgment depth for various rankers.

\begin{table*}[t]
\caption{
Relevance judgment agreement (Cohen's $\kappa$) between TREC assessors and each \ac{LLM}, and Pearson's $\rho$ correlation coefficients between BM25's actual nDCG@10 values and those predicted by \ourmodel integrated with each \ac{LLM} on TREC-DL 19--22.
The best value in each column is marked in \textbf{bold}.
We do not fine-tune Llama-3-70B-Instruct due to budget constraints.
}
\label{tab:finetuning}
\setlength{\tabcolsep}{1mm}
\begin{tabular}{l ll ll ll ll}
\toprule  
& \multicolumn{2}{c}{TREC-DL 19} & \multicolumn{2}{c}{TREC-DL 20} & \multicolumn{2}{c}{TREC-DL 21} & \multicolumn{2}{c}{TREC-DL 22} \\
 \cmidrule(lr){2-3}  \cmidrule(lr){4-5}  \cmidrule(lr){6-7}   \cmidrule(lr){8-9}
 LLM
  & \multicolumn{1}{c}{$\kappa$} & \multicolumn{1}{c}{P-$\rho$}  & \multicolumn{1}{c}{$\kappa$} & \multicolumn{1}{c}{P-$\rho$}   & \multicolumn{1}{c}{$\kappa$} & \multicolumn{1}{c}{P-$\rho$}  & \multicolumn{1}{c}{$\kappa$} & \multicolumn{1}{c}{P-$\rho$} \\
\midrule    
 GPT-3.5 (text-davinci-003)~\citep{faggioli2023perspectives} & \multicolumn{1}{c}{-} & \multicolumn{1}{c}{-} & \multicolumn{1}{c}{-}  & \multicolumn{1}{c}{-}  & 0.260  & \multicolumn{1}{c}{-} & \multicolumn{1}{c}{-}  & \multicolumn{1}{c}{-} \\
  \midrule  
Llama-3.2-1B-Instruct (few-shot) & 0.013  & 0.152  & 0.029  & 0.099  & 0.009  & 0.249 & 0.079 & 0.087 \\
Llama-3.2-3B-Instruct (few-shot) & 0.186  & 0.293  &  0.114  & 0.020  & 0.165  & 0.289 & 0.055 & 0.443 \\
Mistral-7B-Instruct-v0.3 (few-shot) & 0.224  & 0.271  & 0.174  & 0.499  & 0.245 & 0.414 & 0.042 & 0.243 \\
LLaMA-7B (few-shot) & \llap{-}0.001   & \llap{-}0.062  & \llap{-}0.003 & 0.087 &  0.003 & \llap{-}0.002 & \llap{-}0.010 & 0.214 \\
Llama-3-8B (few-shot) & 0.018   & 0.042  & 0.027 & 0.087 & 0.021 & 0.180  & \llap{-}0.035  & 0.087   \\
 Llama-3-8B-Instruct (few-shot)  & 0.315    &  0.510 & 0.227 & 0.372  & 0.238 & 0.462 & 0.049 & 0.388 \\
Mistral-22B-Instruct (few-shot) & 0.281   & 0.412   &  0.238  & 0.535  & 0.276 & 0.528  & 0.083  &  0.473 \\
Llama-3-70B-Instruct (few-shot) & 0.321   & 0.526   & 0.245  & 0.557   & 0.279  & 0.545 & \textbf{0.086}  & 0.483 \\

 \midrule   
Llama-3.2-1B-Instruct (fine-tuned) & 0.351   & 0.610  &  0.211  & 0.596  & 0.197 & 0.570 &  0.042 &  0.428 \\
Llama-3.2-3B-Instruct (fine-tuned) & 0.383  & 0.710  & 0.273  & 0.722  & 0.306 & 0.608 & 0.042 &  0.511  \\
Mistral-7B-Instruct-v0.3 (fine-tuned) & 0.403  & 0.734  & 0.328  & 0.720  & 0.373 & 0.592 & 0.076  & 0.411 \\
 LLaMA-7B (fine-tuned)   &  0.258  &  \textbf{0.715}   & 0.238   & 0.627 &  0.333 & 0.547 & 0.038  & 0.388  \\
 Llama-3-8B (fine-tuned)  &  0.381  &  0.544 & \textbf{0.342}  & 0.681 & 0.347 & 0.612  & 0.082 & 0.568 \\
 Llama-3-8B-Instruct (fine-tuned) & 0.397   & 0.647   & 0.316 & \textbf{0.743}  & \textbf{0.418}  & \textbf{0.699} & 0.066  & \textbf{0.573}   \\ 
Mistral-22B-Instruct (fine-tuned) & \textbf{0.407}  & 0.682  & 0.276  &  0.640  & 0.310 & 0.591 & 0.071 & 0.462  \\
  \bottomrule
\end{tabular}
\end{table*}

\begin{figure*}[t]
    \centering
    \begin{subfigure}{0.495\columnwidth}
        \includegraphics[width=\linewidth]{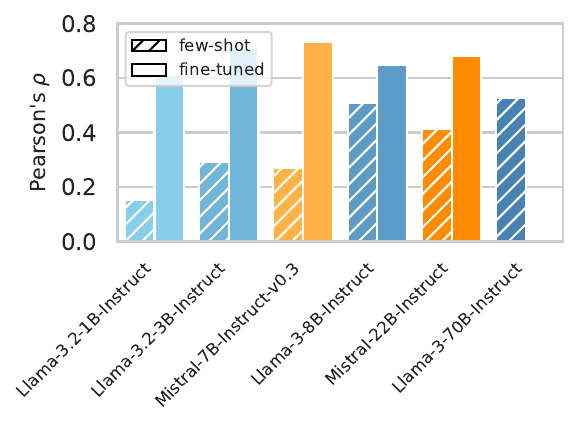}
        \vspace*{-7mm}
        \caption{TREC-DL 19}
        \label{fig:llm_trec-dl19}
    \end{subfigure}
    \begin{subfigure}{0.495\columnwidth}
        \includegraphics[width=\linewidth]{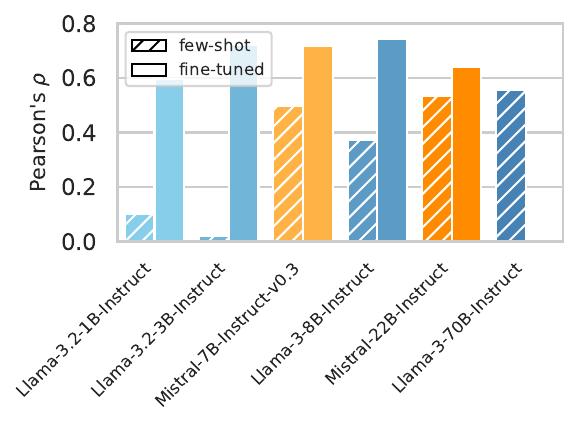}
        \vspace*{-7mm}
        \caption{TREC-DL 20}
        \label{fig:llm_trec-dl20}
    \end{subfigure}
    \begin{subfigure}{0.495\columnwidth}
        \includegraphics[width=\linewidth]{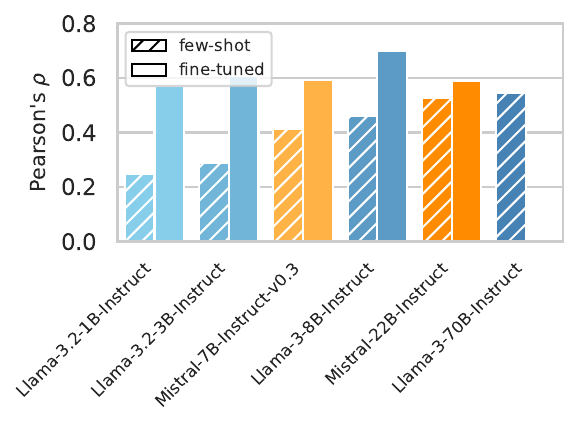}
        \vspace*{-7mm}
        \caption{TREC-DL 21}
        \label{fig:llm_trec-dl21}
    \end{subfigure}
    \begin{subfigure}{0.495\columnwidth}
        \includegraphics[width=\linewidth]{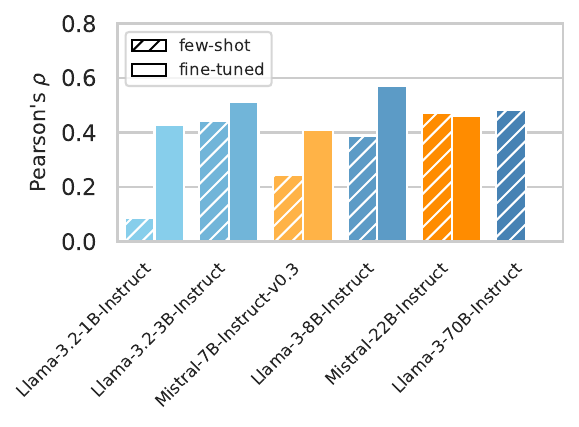}
        \vspace*{-7mm}
        \caption{TREC-DL 22}
        \label{fig:llm_trec-dl22}
    \end{subfigure}
    \caption{
    Pearson's $\rho$ correlation coefficients between BM25' actual nDCG@10 values and those predicted by \ourmodel integrated with various \acp{LLM} with different sizes, under both few-shot and fine-tuned settings, on TREC-DL 19--22.
    From left to right along the x-axis, as the size of the \acp{LLM} increases, the inference efficiency correspondingly decreases.
    To aid visual comparison, \acp{LLM} from the same family share a consistent colour scheme: Llama models are shown in blue, and Mistral models in orange.
    Note that, for simplicity, we only retain the results of \acp{LLM} with ``instruct'' versions.
    We do not fine-tune Llama-3-70B-Instruct due to budget constraints.
    }
    \label{fig:llm}
\end{figure*}

\subsection{Impact of fine-tuning and the choice of LLMs} 
\label{sec:fewshot}

To answer \ref{RQ4}, we analyze the impact of fine-tuning and the choice of \acp{LLM} on the quality of generated relevance judgments and \ac{QPP}.
We evaluate two widely-used families of \acp{LLM}, Llama and Mistral, spanning from 1B to 70B under two settings:
\begin{enumerate*}[label=(\roman*)]
\item trained with \ac{PEFT} on human relevance labels (following the same fine-tuning setup as in Section~\ref{implementation}), and
\item few-shot prompted~(in-context learning)  
\end{enumerate*}.\footnote{We randomly sample human-labeled demonstration examples from the same set used for fine-tuning \acp{LLM}; each example is a triplet (query, passage, relevant/irrelevant); our experiments show that two examples work best, one with relevant passages and one with irrelevant passages.} 
For the Llama family, besides LLaMA-7B~\citep{touvron2023llama}, our evaluation includes Llama-3.2-1B-Instruct, Llama-3.2-3B-Instruct, Llama-3-8B, Llama-3-8B-Instruct, and Llama-3-70B-Instruct.
For the Mistral family, we focus on Mistral-7B-Instruct-v0.3 and Mistral-22B-Instruct (a.k.a. Mistral-Small-Instruct-2409).

We do not report the results for a zero-shot setting because our preliminary experiments show that prompting these \acp{LLM} in a zero-shot way yields pretty poor performance.
Note that we do not fine-tune Llama-3-70B-Instruct due to budget constraints.

%

To evaluate the performance of judging relevance, we compute Cohen's $\kappa$ metric to measure the agreement between relevance judgments made by the TREC assessors (i.e., relevance judgments in the \textit{qrels}) and relevance judgments automatically generated by a fine-tuned or few-shot \ac{LLM}, on TREC-DL 19--22.
\citet{faggioli2023perspectives} reported the relevance judgment agreement in terms of Cohen's $\kappa$ between TREC assessors and GPT-3.5 (text-davinci-003) on TREC-DL 21; we also consider their Cohen's $\kappa$ value for  comparison.
To evaluate \ac{QPP} quality, we compute the Pearson's $\rho$ correlation coefficients between BM25's actual nDCG@10 values and those predicted by \ourmodel using relevance judgments generated by an \ac{LLM}, on TREC-DL 19--22.\footnote{We do not report the Pearson's $\rho$ correlation for GPT-3.5~(text-davinci-003) because the relevance judgments generated by \citet{faggioli2023perspectives} are not available to us.}
The judging depth is set to 1000 in a ranked list.
We show the results in Table~\ref{tab:finetuning} and Figure~\ref{fig:llm}.

We have three observations.
First, fine-tuning generally markedly improves the quality of relevance judgment generation and \ac{QPP}, particularly for \acp{LLM} with sizes ranging from 1 billion to 8 billion parameters.
Specifically, almost all of fine-tuned \acp{LLM} exhibit improved relevance judgment agreement with the TREC assessors on TREC-DL 19--22.
After fine-tuning, LLaMA-7B and Llama-3-8B achieve ``fair'' agreement with the TREC assessors on TREC-DL 19, 20 and 21,\footnote{Note that unlike the qrels files for TREC-DL 19, 20, and 21 which are fully manually annotated, the qrels file for TREC-DL 22 is constructed by first detecting near-duplicate items and manually judging only one representative item from each near-duplicate cluster for a given query~\citep{2021craswell}; this difference may result in variation in Cohen's $\kappa$ values of LLMs across TREC-DL 19, 20, 21, and TREC-DL 22.} Llama-3-8B-Instruct (fine-tuned) even achieves ``moderate'' agreement on TREC-DL 21 (a Cohen's $\kappa$ value of 0.418).
All fine-tuned \acp{LLM} (except for Llama-3.2-1B-Instruct) exhibit a higher Cohen's $\kappa$ value than the commercial \ac{LLM}, GPT-3.5 (text-davinci-003).
All fine-tuned \acp{LLM} (except for Mistral-22B-Instruct on TREC-DL 22) surpass their corresponding few-shot counterpart on all datasets in terms of Pearson's $\rho$.
This reveals that fine-tuning is an effective way to improve the quality of \acp{LLM} in generating relevance judgments, which finally translates to better \ac{QPP} quality.

Second, larger \acp{LLM} with over 22 billion parameters demonstrate significantly greater effectiveness than their smaller counterparts in the few-shot setting. 
Specifically, in this setting, Llama-3-70B-Instruct achieves the best overall performance.
Mistral-22B-Instruct consistently outperforms Mistral-7B-Instruct-v0.3 across all datasets.
However, a fine-tuned Llama model with only 3 billion parameters markedly outperforms both of these larger few-shot \acp{LLM} across all datasets.

Third, Instruction-tuned \acp{LLM} generally perform better than their standard counterparts. 
Llama-3-8B-Instruct further enhances relevance judgment generation and \ac{QPP} quality over both Llama-3-8B and LLaMA-7B across most cases. 
Notably, Llama-3-8B-Instruct (few-shot) even performs better than or equally as well as LLaMA-7B (fine-tuned) on TREC-DL 19 and 22.
This finding implies that with a more effective \ac{LLM} \ourmodel has the potential to achieve improved \ac{QPP} performance.

The above observations provide insights into the minimum requirements needed to achieve reliable \ac{QPP} quality.
Our findings show that a fine-tuned 3B model (Llama-3.2-3B-Instruct) offers the best trade-off between \ac{QPP} quality and computational overhead: it not only markedly outperforms few-shot 70B \acp{LLM}, but also delivers \ac{QPP} quality comparable to that of fine-tuned 7/8B \acp{LLM}.




\begin{figure}[t]
    \centering
    \begin{subfigure}{0.495\columnwidth}
        \includegraphics[width=\linewidth]{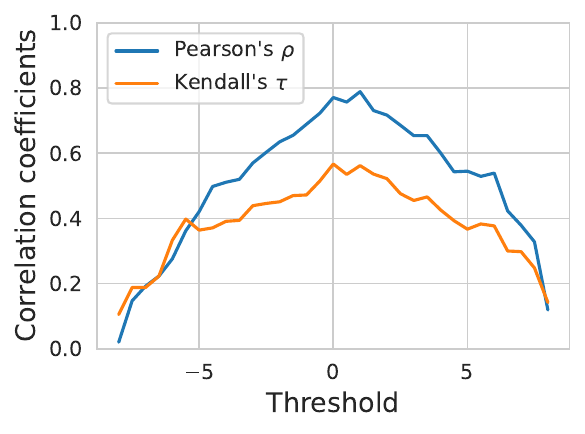}
        \caption{BM25 on TREC-DL 19}
        \label{rankllama-ndcg10-dl19-bm25}
    \end{subfigure}
    \begin{subfigure}{0.495\columnwidth}
        \includegraphics[width=\linewidth]{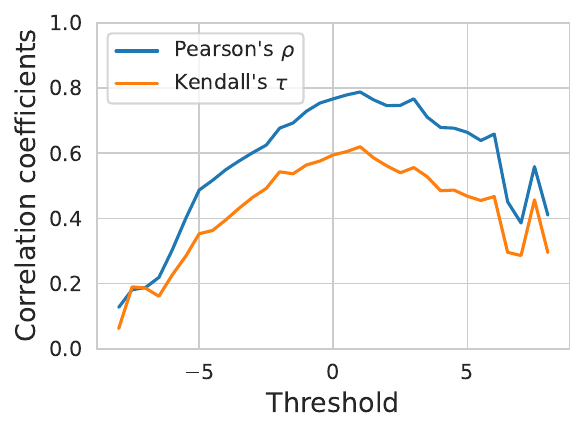}
        \caption{BM25 on TREC-DL 20}
        \label{rankllama-ndcg10-dl20-bm25}
    \end{subfigure}
    \caption{
    \ac{QPP} quality of \ourmodel integrated with RankLLaMA~\citep{ma2023fine} in predicting nDCG@10 values for BM25, w.r.t. threshold values ranged from -8 to 8, on TREC-DL 19 and 20. 
    An item is predicted as ``relevant'' if its re-ranking score meets or exceeds a given threshold value.
    }
    \label{fig:reranker}
\end{figure}

\subsection{Integrating \ourmodel with an LLM-based re-ranker}
\label{sec:reranker}

To show \ourmodel's compatibility with other types of relevance prediction methods instead of directly asking an \ac{LLM} to explicitly generate explicit relevance judgments, we adapt a state-of-the-art pointwise \ac{LLM}-based re-ranker, RankLLaMA~\citep{ma2023fine}, into a relevance judgment generator, and then integrate \ourmodel with the adapted RankLLaMA.
Specifically, we translate a re-ranking score into a relevance judgment by applying a threshold: an item is deemed as ``relevant'' if its re-ranking score meets or exceeds a given threshold value.
We analyze Pearson's $\rho$ and Kendall's $\tau$ correlation coefficients between BM25’s actual nDCG@10 values and those predicted by \ourmodel integrated with RankLLaMA w.r.t.\ different threshold values on TREC-DL 19 and 20.
We employ RankLLaMA (7B) from Tevatron.\footnote{\url{https://github.com/texttron/tevatron/tree/main/examples/rankllama}}
RankLLaMA's re-ranking scores for BM25 range from -12.93 to 89.90 for TREC-DL 19 and from -14.38 to 8.82 for TREC-DL 20. 
Thresholds are set at intervals of 0.5. 
The judging depth is set to 1,000 in a ranked list.

We report the results in Figure~\ref{fig:reranker}.
We find that RankLLaMA achieves the highest \ac{QPP} quality on both datasets when the threshold is 1.
At this particular threshold, RankLLaMA achieves high Pearson's $\rho$ values of 0.789 and 0.788 on TREC-DL 19 and 20, respectively.
These values exceed those of fine-tuned LLaMA-7B, which achieves Pearson's $\rho$ values of 0.715 and 0.627 on TREC-DL 19 and 20, respectively, as well as Llama-3-8B-Instruct, which achieves Pearson's $\rho$ values of 0.647 and 0.743 on TREC-DL 19 and 20, respectively (see Figure~\ref{tab:finetuning}).\footnote{Note that the comparison is not fair because 
\begin{enumerate*}[label=(\roman*)]
\item \acp{LLM} and all other supervised \ac{QPP} methods used in this paper are trained on the development set of MS MARCO V1, while RankLLaMA~\citep{ma2023fine} was trained on the training set of MS MARCO V1, which is much larger. 
\item We employ the official version of MS MARCO V1, while RankLLaMA~\citep{ma2023fine} uses the Tevatron version of MS MARCO V1, where passages are enriched with document titles; \citet{lassance2023tale} reveal that incorporating titles leads to enhanced ranking performance.
\end{enumerate*}}
This means that a state-of-the-art pointwise \ac{LLM}-based re-ranker can be adapted into an effective relevance judgment generator.
The high \ac{QPP} quality achieved by RankLLaMA demonstrates \ourmodel's compatibility with other types of relevance prediction methods besides directly using \acp{LLM} as relevance judgment generators (i.e., asking an \ac{LLM} to explicitly generate explicit relevance judgments).

However, compared to directly regarding an \ac{LLM} as a relevance judgment generator, adapting an \ac{LLM}-based re-ranker into a relevance judgment generator requires tuning an appropriate threshold.
As demonstrated, re-ranking scores are not normalized and their ranges vary across datasets.
Directly using a re-ranker as a relevance judgment generator can cause issues in real-world scenarios.
Extra calibration work for re-ranking scores might be necessary.

\begin{table*}[!t]
\caption{
Correlation coefficients (Pearson's $\rho$ and Kendall's $\tau$) between actual retrieval quality, in terms of nDCG@3, of ConvDR~\citep{yu2021few} and its performance predicted by \ourmodel/baselines, on CAsT-19 and 20.
Following \citet{meng2023query}, for \ac{QPP} methods requiring queries as input, we feed them with either T5-generated or human-written query rewrites.
$^*$ indicates statistically significant correlation coefficients ($p$-value $< 0.05$). 
$^\dagger$ indicates the statistically significant improvement of \ourmodel ($n$=200) compared to all the baselines (paired $t$-test; $p$-value $< 0.001$ with Bonferroni correction for multiple testing).
The best value in each column is marked in \textbf{bold}.
$n$ denotes \ourmodel's judgment depth in a ranked list.
}
\label{tab:cs}
\setlength{\tabcolsep}{1mm}
\begin{tabular}{l ll ll ll ll}
\toprule  
& \multicolumn{8}{c}{Ranker: ConvDR} \\
 \cmidrule(lr){2-9}
 & \multicolumn{4}{c}{T5-generated query rewrites} & \multicolumn{4}{c}{Human-written query rewrites} \\
 \cmidrule(lr){2-5}   \cmidrule(lr){6-9} 
& \multicolumn{2}{c}{CAsT-19} & \multicolumn{2}{c}{CAsT-20}  & \multicolumn{2}{c}{CAsT-19} & \multicolumn{2}{c}{CAsT-20} \\
 \cmidrule(lr){2-3}  \cmidrule(lr){4-5}  \cmidrule(lr){6-7}   \cmidrule(lr){8-9}
 QPP method
  & \multicolumn{1}{c}{P-$\rho$} & \multicolumn{1}{c}{K-$\tau$}  & \multicolumn{1}{c}{P-$\rho$} & \multicolumn{1}{c}{K-$\tau$}   & \multicolumn{1}{c}{P-$\rho$} & \multicolumn{1}{c}{K-$\tau$}  & \multicolumn{1}{c}{P-$\rho$} & \multicolumn{1}{c}{K-$\tau$} \\
\midrule        
Clarity & 0.257$^*$  & 0.176$^*$   & 0.126  & 0.088     &  0.257$^*$ &  0.176$^*$ & 0.126  &  0.088 \\
 WIG &  0.387$^*$ & 0.274$^*$  & 0.377$^*$   & 0.277$^*$   &  0.412$^*$     & 0.285$^*$  &  0.384$^*$   &  0.264$^*$  \\
 NQC  & 0.431$^*$  & 0.307$^*$  &  0.339$^*$   &  0.261$^*$  & 0.431$^*$  & 0.307$^*$  &  0.339$^*$   &  0.261$^*$ \\
 $\sigma_{max}$  &  0.378$^*$  & 0.267$^*$  & 0.282$^*$ & 0.219$^*$   &  0.378$^*$  & 0.267$^*$  & 0.282$^*$ & 0.219$^*$   \\
 n($\sigma_{x\%}$)   &  0.187$^*$ & 0.175$^*$  &  0.199$^*$  &  0.168$^*$ & 0.216$^*$   &  0.196$^*$   &  0.201$^*$   & 0.156$^*$    \\
 SMV & 0.386$^*$   & 0.285$^*$   & 0.275$^*$   & 0.216$^*$  & 0.386$^*$   & 0.285$^*$   & 0.275$^*$   & 0.216$^*$   \\
UEF(NQC)  & 0.435$^*$  & 0.312$^*$  &  0.343$^*$   &  0.265$^*$  & 0.427$^*$  & 0.310$^*$  &  0.341$^*$   &  0.263$^*$ \\
RLS(NQC)  & 0.429$^*$  & 0.311$^*$  &  0.337$^*$   &  0.267$^*$  & 0.413$^*$  & 0.308$^*$  &  0.342$^*$   &  0.259$^*$ \\
\ac{QPP-PRP} & 0.350$^*$  & 0.270$^*$  &  0.280$^*$   &  0.210$^*$  & 0.345$^*$  & 0.265$^*$  &  0.275$^*$   &  0.205$^*$ \\
 \midrule 
NQA-QPP  & 0.175  & 0.115  & 0.082 & 0.075  & 0.142  & 0.091  & 0.065  & 0.058 \\
BERTQPP    & 0.243$^*$ & 0.170$^*$ & 0.236$^*$ & 0.185$^*$  & 0.256$^*$ & 0.172$^*$  &  0.262$^*$ & 0.209$^*$ \\
qppBERT-PL  & 0.203$^*$ & 0.169$^*$  & 0.181$^*$ & 0.165$^*$ & 0.105 & 0.090  & 0.166$^*$ & 0.161$^*$ \\ 
\ac{M-QPPF} & 0.242$^*$ & 0.174$^*$  & 0.285$^*$ & 0.219$^*$ & 0.262$^*$ & 0.190$^*$  & 0.313$^*$ & 0.254$^*$ \\ 
\midrule  
\ourmodel ($n=200$)  & \textbf{0.623}$^\dagger$$^*$  & \textbf{0.505}$^\dagger$$^*$ & 0.484$^\dagger$$^*$ & 0.395$^\dagger$$^*$  & \textbf{0.645}$^\dagger$$^*$ & \textbf{0.529}$^\dagger$$^*$ & 0.678$^\dagger$$^*$ & 0.551$^\dagger$$^*$ \\
\midrule 
\ourmodel ($n=10$)  & 
0.617$^*$  & 0.504$^*$ & 0.471$^*$ & 0.388$^*$  & 0.619$^*$  & 0.506$^*$ & 0.659 & 0.534 \\
\ourmodel ($n=100$)  &  \textbf{0.623}$^*$  & \textbf{0.505}$^*$ &  0.485$^*$ & 0.396$^*$ &  0.644$^*$ &  \textbf{0.529}$^*$ & 0.675$^*$ & 0.547$^*$ \\
\ourmodel ($n=1,000$)  & \textbf{0.623}$^*$ & \textbf{0.505}$^*$ & \textbf{0.487}$^*$ & \textbf{0.398}$^*$  &  \textbf{0.645}$^*$ & \textbf{0.529}$^*$ & \textbf{0.684}$^*$ & \textbf{0.556}$^*$\\
\bottomrule
\end{tabular}
\end{table*}

\subsection{Generalization to conversational search}
\label{sec:cs}

To assess the generalizability of \ourmodel to new domains, we apply it to the conversational search scenario~\citep{mo2023convgqr,mo2023learning,mo2024survey,mo2024chiq,mo2025conversational,meng2025bridging} in a zero-shot manner.
Specifically, we evaluate \ourmodel and other baselines on predicting the performance of ConvDR~\citep{yu2021few}, a widely used conversational dense retriever.
Given the findings in Section~\ref{sec:fewshot}, which show that the fine-tuned Llama-3.2-3B-Instruct model achieves high relevance prediction quality at low inference cost, we equip \ourmodel with this model fine-tuned on MS MARCO V1 for relevance prediction.
For all supervised \ac{QPP} baselines, we directly use their checkpoints trained on MS MARCO V1 for assessing ANCE (see Section \ref{subsec:recall}).
Because a user query in a conversation depends on the conversational context, i.e., a query may contain omissions, coreferences, or ambiguities, it is hard for existing \ac{QPP} methods to capture users' information need from such context-dependent queries.
Therefore, we follow \citet{meng2023query} to provide \ac{QPP} methods (including \ourmodel) with self-contained query rewrites as input. 
These rewrites are either generated by the T5 query generator\footnote{\url{https://huggingface.co/castorini/t5-base-canard}} or written by humans.
Table~\ref{tab:cs} presents the performance of \ourmodel along with all the baselines on assessing ConvDR~\citep{yu2021few} in terms of nDCG@3 on the CAsT-19~\citep{dalton2020cast} and 20~\citep{Dalton2020CAsT2T} datasets. 
Note that nDCG@3 is the primary evaluation metric officially adopted by TREC CAsT~\citep{dalton2020cast}.
We have two main observations.

First, \ourmodel significantly outperforms all \ac{QPP} baselines on both datasets when provided with either type of query input. 
This demonstrates the ability of \ourmodel to generalize effectively to the conversational search domain.
Second, \ourmodel achieves higher performance when provided with human-written query rewrites compared to T5-generated rewrites on both datasets. 
This finding highlights the critical role of high-quality query rewrites in effectively adapting \ac{QPP} methods to conversational search scenarios; this finding also aligns with prior research \citep{meng2023query}.

\subsection{\ourmodel's interpretability}
\label{sec:interpretability}

\begin{figure}[t]
    \centering
    \begin{subfigure}{0.495\columnwidth}
        \includegraphics[width=\linewidth]{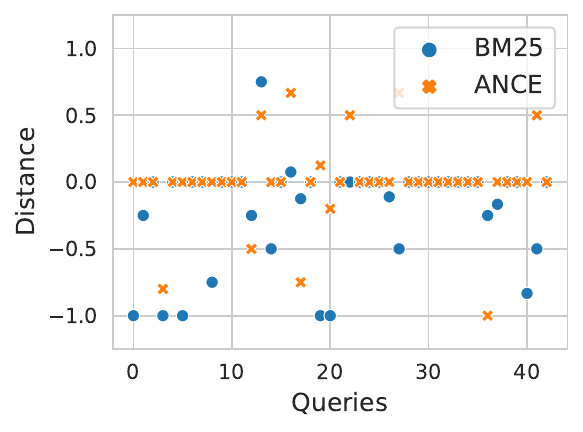}
        \caption{TREC-DL 19}
        \label{fig:error-dl19}
    \end{subfigure}
    \begin{subfigure}{0.495\columnwidth}
        \includegraphics[width=\linewidth]{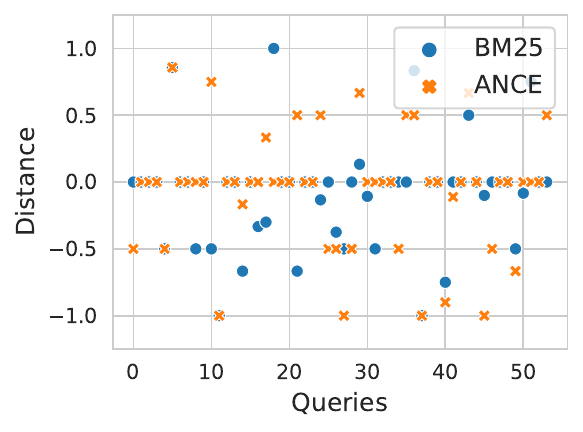}
        \caption{TREC-DL 20}
        \label{fig:error-dl20}
    \end{subfigure}
    \caption{
    The \ac{QPP} errors of \ourmodel integrated with LLaMA-7B in predicting the performance of BM25 and ANCE in terms of RR@10 on TREC-DL 19 and 20. 
    The distance is defined as ``predicted RR@10 minus actual RR@10.''
    The closer a query point is to 0 on the Y-axis, the more accurately \ourmodel predicts its difficulty.
    }
    \label{fig:error}
\end{figure}

\begin{table}[t!]
\centering
\caption{
Confusion matrices comparing relevance judgments made by TREC assessors and \ourmodel integrated with LLaMA-7B on TREC-DL 19 and 20.
}
\label{tab:confusion} 
\begin{tabular}{l cccc}
\toprule
\multirow{2}{*}{\ourmodel} &\multicolumn{2}{c}{TREC-DL 19 assessors} &\multicolumn{2}{c}{TREC-DL 20 assessors} \\
\cmidrule(lr){2-3} \cmidrule(lr){4-5} 
 & Relevant  & Irrelevant   & Relevant   & Irrelevant  \\
\midrule
Relevant  & \phantom{0}752 & \phantom{0}553  & \phantom{0}486 & \phantom{0}763 \\
Irrelevant & 1,749 & 6,206  & 1,180 & 8,957 \\
\bottomrule
\end{tabular}
\end{table}

\begin{table}[t!]
\centering
{
\caption{
Performance of each class for \ourmodel with LLaMA-7B on the TREC-DL 2019 and 2020 qrels.}
\label{tab:metrics}
\begin{tabular}{l ccc ccc}
\toprule
 & \multicolumn{3}{c}{TREC-DL 19 } &\multicolumn{3}{c}{TREC-DL 20} \\
\cmidrule(lr){2-4} \cmidrule(lr){5-7} 
Class
     & Precision & Recall & F1  & Precision & Recall & F1 \\ 
    \midrule
Relevant   & 0.576   & 0.301  & 0.395   & 0.389              & 0.292           & 0.333        \\
Irrelevant         & 0.780              & 0.918           & 0.844  & 0.884              & 0.922           & 0.902           \\ 
\bottomrule
\end{tabular}
} 
\end{table}

\begin{table}[h!]
\centering
\caption{
Retrieval quality, in terms of RR@10, predicted by various \ac{QPP} methods for the BM25 retrieval of the query ``who is Robert Gra'' (query ID 1037798) on TREC-DL 19.
}
\label{tab:case1}
\begin{tabular}{l c} 
\toprule
\textbf{QPP Methods} & \textbf{Retrieval quality} \\ \hline
WIG & 2.427 \\ 
NQC & 0.106 \\ 
BERTQPP & 0.172 \\ 
qppBERT-PL & 0.368 \\ 
QPP-GenRE & 0.500 \\ 
 \midrule 
Ground-truth RR@10 & 1.000 \\
\bottomrule
\end{tabular}
\end{table}

\begin{table}[h!]
\centering
\caption{
Ranked list returned by BM25 for the query ``who is Robert Gra'' (query ID 1037798), human-labeled relevance judgments and ones predicted by \ourmodel with LLaMA-7B on TREC-DL 19.}
\label{tab:case2}
\begin{tabular}{c p{8.5cm} ll }
\toprule
Rank & Passage & Human & \ourmodel \\ \hline

\centering 1 & Captain Robert Gray, May 1972. Discovering the Columbia River, May 1792 ... The Columbia River was given the name it bears today in May 1792...
& Relevant & Irrelevant \\ \midrule 

\centering 2 & Robert Gray. A surprise came on the Democratic side in the race for Mississippi Governor. Robert Gray, a retired firefighter and truck driver...  
& Irrelevant & Relevant \\ \midrule 

\centering 3 & Team Mississippi Robert Gray For Governor Official Page. Robert Gray never would have made it without God...
& Irrelevant & Irrelevant \\ \midrule 

\centering 4 & I'm not a politician, said Gray in a Wednesday interview. I'm not a person who really wanted to run for Governor. Robert Gray is a 46-year-old truck driver...
& Irrelevant & Relevant \\ 
\bottomrule
\end{tabular}
\end{table}

As \ourmodel computes \ac{QPP} based on generated relevance judgments, we analyze \ac{QPP} errors from the perspective of relevance judgment generation.
Figure~\ref{fig:error} shows the \ac{QPP} errors of \ourmodel integrated with LLaMA-7B in predicting the performance of BM25 and ANCE in terms of RR@10 on TREC-DL 19 and 20; the error is defined as the distance between the RR@10 values predicted by \ourmodel and actual RR@10 values, namely ``predicted RR@10 minus actual RR@10.''
We find that most RR@10 values predicted by \ourmodel tend to be smaller than the actual RR@10 values, indicating that \ourmodel performs less effectively in identifying relevant items than irrelevant ones in the top of the ranked list.
Table~\ref{tab:confusion} shows the confusion matrices that compare relevance judgments made by TREC assessors~(i.e., relevance judgments in \textit{qrels}) and \ourmodel integrated with LLaMA-7B on TREC-DL 19 and 20.
Table~\ref{tab:metrics} provides a detailed breakdown of \ourmodel's prediction performance for each class, including metrics such as Precision, Recall, and F1 score.
We find that \ourmodel tends to wrongly predict some relevant items as irrelevant (false negatives), which provides a further interpretation of the \ac{QPP} errors we found above.
Therefore, reducing false negatives in generating relevance judgments is a potential way to improve the \ac{QPP} quality of \ourmodel.
We leave this exploration for future work.

To show the superior interpretability of \ourmodel compared to other baselines, we provide a case study shown in Tables~\ref{tab:case1} and Tables~\ref{tab:case2}.
Table~\ref{tab:case1} lists the predicted or ground-truth retrieval quality in terms of RR@10 of BM25 for the query ``who is Robert Gra'' on TREC-DL 2019. 
The predictions are made using widely-used unsupervised \ac{QPP} methods (WIG, NQC), supervised \ac{QPP} methods (BERTQPP, qppBERT-PL), and \ourmodel.

We observe that the ground-truth RR@10 score for the query is 1, while the score predicted by \ourmodel is 0.5. 
\ourmodel infers RR@10 directly from the predicted relevance judgments for items in BM25's ranked list, we can conclude that while the top-ranked item is actually relevant to the query, \ourmodel mistakenly classified it as irrelevant. 
Table~\ref{tab:case2} provides supporting evidence by displaying the relevance judgments assigned by human annotators and \ourmodel for each item in BM25's ranked list.
We found that \ourmodel fails to identify the top-ranked item as relevant.
Specifically, \ourmodel does not identify the key part ``Captain Robert Gray'' in this item.
This suggests that we could potentially improve \ourmodel's performance by further fine-tuning it to predict relevance on query--item pairs specifically related to influential historical figures.

However, all baselines lack interpretability compared to \ourmodel.
WIG and NQC do not directly predict values for a specific \ac{IR} metric, and their scores are difficult to interpret in isolation without comparing them to the scores for other queries.
BERTQPP is trained to predict RR@10, but in this case, it returns a score of 0.3, indicating inaccurate performance prediction. 
Unfortunately, BERTQPP does not provide any intermediate outputs to help understand why it made this error or how its performance can be improved, limiting its interpretability and actionable insights.
qppBERT-PL first predicts the number of relevant items in each chunk of the ranked list and then aggregates these numbers into an overall score.
While it is possible to check the intermediate predictions of the number of relevant items per chunk, this information is too coarse to provide detailed insights. 
In contrast, \ourmodel predicts the relevance of each individual item, offering more granular and informative insights.

\subsection{Computational cost analysis}
\label{sec:cost}
%
\begin{table}[!t]
\centering
\caption{
 Inference efficiency of supervised \ac{QPP} baselines and \ourmodel integrated with LLaMA-7B on TREC-DL 19 to predict 1--4 different \ac{IR} metrics. 
 $n$ denotes \ourmodel's judgment depth in a ranked list.
 Cases with higher latency than \ourmodel ($n=10$) are \underline{underlined}. 
}
\label{tab:cost}
\begin{tabular}{l cccc} 
\toprule
  & \multicolumn{4}{c}{Inference latency per query (ms)}     \\
 \cmidrule{2-5}
 \ac{QPP} Method
                            & 1 & 2 & 3 & 4 \\
\midrule

NQA-QPP                     & \phantom{0}118.40  & \phantom{0}236.80 & \phantom{0}355.20 & \phantom{0}\underline{473.60} \\
BERTQPP                     & \phantom{00}30.29   & \phantom{00}60.58 & \phantom{00}90.87 & \phantom{0}121.16  \\
qppBERT-PL                  & \phantom{0}316.80  & \phantom{0}316.80 & \phantom{0}316.80 & 316.80 \\
\ac{M-QPPF}                 & \phantom{0}289.27  & \phantom{0}\underline{578.54} & \phantom{0}\underline{867.81} & \underline{1157.08} \\
\midrule
\ourmodel ($n=10$)          & \phantom{0}452.60 & \phantom{0}452.60 & \phantom{0}452.60 & \phantom{0}452.60   \\
\ourmodel ($n=100$)         & 1,566.25 & 1,566.25 & 1,566.25 & 1,566.25 \\
\ourmodel ($n=200$)         & 2,845.43 & 2,845.43 & 2,845.43 & 2,845.43 \\
\bottomrule
\end{tabular}
\end{table}

Table~\ref{tab:cost} shows the online \ac{QPP} latency of \ourmodel integrated with LLaMA-7B and other BERT-based supervised \ac{QPP} baselines, on TREC-DL 19, on a single NVIDIA A100 GPU.
We compute the inference latency when queries are processed individually.
For \ourmodel, we consider judging depths at 10, 100, and 200; \ourmodel can use batch acceleration for judging items for the same query because each item in a ranked list for a query is independent of each other.\footnote{qppBERT-PL first splits a ranked list with 100 items into 25 chunks and then predicts the number of relevant items in each chunk. For a fair comparison, we put 25 chunks into one batch for acceleration.}
Although \ourmodel is more expensive than all baselines when predicting one measure due to the much larger parameter size of LLaMA-7B compared to BERT,  
\ourmodel has lower latency compared to some baselines when predicting multiple \ac{IR} evaluation measures because multiple measures can be derived from the same set of relevance judgments at no additional cost.
E.g., while \ourmodel is 56\% more expensive than M-QPPF for predicting one measure, it becomes more efficient when predicting 2 or more metrics than M-QPPF. 
Nevertheless, we acknowledge that \ourmodel has higher computational costs than supervised \ac{QPP} methods when predicting a single measure.
Conversely, regression-based \ac{QPP} baselines (NQA-QPP, BERTQPP and \ac{M-QPPF}) need to train separate models for different \ac{IR} evaluation metrics.
Although qppBERT-PL is not optimized to learn to output one specific \ac{IR} evaluation measure, qppBERT-PL does not achieve a promising \ac{QPP} quality (see Sections~\ref{subsec:precision} and \ref{subsec:recall}).

We argue that \ourmodel's latency is still much smaller than some high-performing \ac{LLM}-based re-rankers.
E.g., \citet{sun2023chatgpt} show that a GPT-4-based listwise re-ranker needs 10 API calls (one call takes 3,200ms) to re-rank 100 items for a query, resulting in 32,000ms in total, which is around 20 times worse than \ourmodel's latency with a judging depth of 100.
\ourmodel can well fit some knowledge-intensive professional search scenarios where \ac{QPP} quality is prioritized or users may have a higher tolerance level for latency.
Besides using \ac{QPP} online, \ac{QPP} can also be used to analyze a search system's performance in an offline setting~\citep{faggioli2023geometric}.

\begin{figure*}[t]
    \centering
    \begin{subfigure}{0.495\columnwidth}
    
        \includegraphics[width=\linewidth]{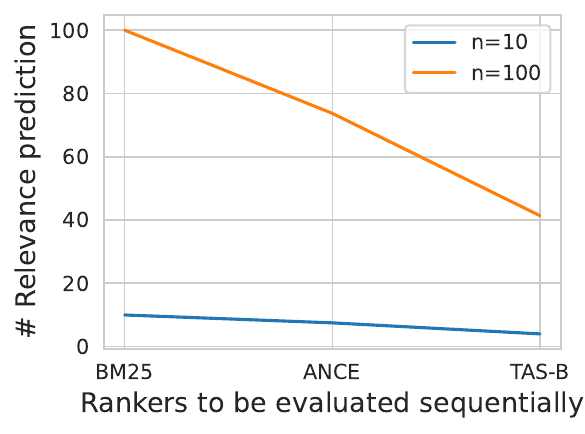}
        \vspace*{-7mm}
        \caption{TREC-DL 19}
        \label{fig:cache_d19}
    \end{subfigure}
    \begin{subfigure}{0.495\columnwidth}
        \includegraphics[width=\linewidth]{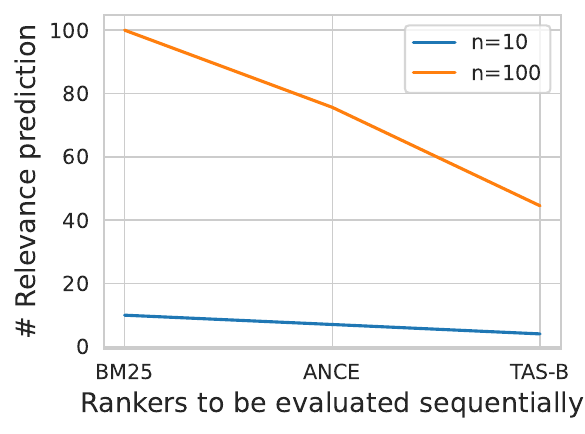}
        \vspace*{-7mm}
        \caption{TREC-DL 20}
        \label{fig:cache_d20}
    \end{subfigure}
    \caption{
    The average number of actual relevance predictions required for sequentially predicting the performance of BM25, ANCE, and TAS-B  using the \textit{relevance judgment caching mechanism}, with a judging depth of $n$ equal to 10 or 100, on TREC-DL 19 and 20.
    }
    \label{fig:cache}
\end{figure*}

Lastly, in order to enhance the efficiency of \ourmodel, we propose a \textit{relevance judgment caching mechanism}.
It reuses previously predicted relevance judgments for the same query when predicting the performance of new rankers.
As a result, this mechanism helps conserve computational resources by eliminating the need to recompute relevance judgments that are shared among multiple rankers.
Figure \ref{fig:cache} shows the number of actual relevance predictions required for sequentially predicting the performance of BM25, ANCE, and TAS-B using the \textit{relevance judgment caching mechanism}, with a judging depth of $n$ equal to 10 or 100, on TREC-DL 19 and 20.
We found that our proposed relevance judgment caching mechanism can reduce the number of \ac{LLM} calls for relevance prediction by approximately 30\%.
For instance, on TREC-DL 19, with a judging depth of 10, the caching mechanism results in 21.15 \ac{LLM} calls on average when sequentially predicting the performance of the three rankers (10 for BM25, 7.06 for ANCE, and 4.09 for TAS-B). 
In contrast, without using this mechanism, 30 \ac{LLM} calls would be required (3 $\times$ 10).

\section{Conclusions \& Future Work}
We have proposed a new \ac{QPP} framework, \ourmodel, which models \ac{QPP} from the perspective of predicting \ac{IR} evaluation measures based on automatically generated relevance judgments.
We have devised an approximation strategy for predicting an \ac{IR} evaluation measure considering recall, which only judges a limited number of items in a given ranked list for a query, to avoid the cost of traversing the entire corpus to find all relevant items; the approximation strategy also enables us to study into the impact of various judging depths on \ac{QPP} quality.
%
We have explored using open-source \acp{LLM} for generating relevance judgments, to ensure scientific reproducibility. 
In addition, we have examined training open-source \acp{LLM} with \acf{PEFT} on human-labeled relevance judgments, to improve the quality of relevance judgment generation and \ac{QPP}. 

\paragraph{Main findings}
Experiments on datasets from the TREC-DL 19--22 tracks demonstrate that \ourmodel significantly surpasses existing \ac{QPP} approaches, achieving state-of-the-art \ac{QPP} quality in assessing lexical and neural rankers for either a precision-oriented \ac{IR} metric or an \ac{IR} metric considering recall.
Moreover, we have shown that fine-tuning open-source LLMs on human-labeled relevance judgments is crucial for obtaining reliable relevance prediction and QPP results.
Fine-tuning much smaller \acp{LLM} for relevance judgment prediction can yield more effective relevance prediction and \ac{QPP} than few-shot prompting with much larger models.
In particular, a fine-tuned 3B model (Llama-3.2-3B-Instruct) offers the best trade-off between \ac{QPP} quality and computational efficiency: it significantly outperforms 70B few-shot models and delivers performance comparable to fine-tuned 7B and 8B models.
It implies that fine-tuning can offer strong performance at low inference costs.
Furthermore, \ourmodel has the potential to conduct \ac{QPP} more accurately when integrated with a more effective \ac{LLM}, has a good compatibility with other types of relevance prediction methods (e.g., an \acp{LLM}-based re-ranker). 
Additionally, We have demonstrated that \ourmodel has great generalizability to the conversational search scenario.
We have shown that \ourmodel exhibits good interpretability.
Finally, we have found that our proposed \textit{relevance judgment caching mechanism } can reduce LLM calls for relevance prediction by about 30\%.

\paragraph{Broader implications} 
\ourmodel has the potential to facilitate the practical use of \ac{QPP}.
The limited accuracy and interpretability of current \ac{QPP} methods make them difficult to use in practical applications~\citep{arabzadeh2024query}.
However, \ourmodel demonstrates significantly improved \ac{QPP} accuracy and better interpretability, enhancing the reliability of \ac{QPP} results and potentially facilitating the practical use of \ac{QPP}.
Especially, \ourmodel has the potential to benefit some knowledge-intensive professional search scenarios.
In such scenarios, accurate \ac{QPP} is prioritized, interpretable \ac{QPP} results are needed, and users may have a higher tolerance level for latency.
\ourmodel also has the potential for practical application in commercial search engines: commercial search engines receive many frequent and repeated queries, and \ourmodel can improve \ac{QPP} efficiency by reusing stored relevance judgments for repeated query-item pairs and only generating relevance judgments for new query-item pairs.
Moreover, \ourmodel can be used to analyze the ranking quality of a search system in an purely offline setting~\citep{faggioli2023geometric}, where latency is not necessarily an issue.

\paragraph{Limitations and future work}
First, we only consider predicting the ranking quality of widely-used lexical and dense retrievers, and have not investigated \ourmodel's bias towards \acp{LLM}-based rankers~\citep{ma2023fine}.
Given that \ourmodel is based on \ac{LLM}-based relevance predictors, it would be particularly interesting to explore \ourmodel's potential biases when it predicts the ranking quality of \ac{LLM}-based rankers.
%

Second, \ourmodel is a \ac{QPP} framework that can be integrated with various relevance prediction approaches.
We show the success of \ourmodel equipped with various open-source \acp{LLM} as well as a state-of-the-art pointwise \ac{LLM}-based re-ranker, RankLLaMA~\citep{ma2023fine}.
Exploring various \acp{LLM} to find the optimal one for relevance prediction is beyond the scope of our work.
However, in future, we believe it is valuable to investigate \ourmodel's performance integrated with other open-source \acp{LLM} as relevance judgment generators.
It is also interesting to adapt pairwise or listwise \ac{LLM}-based re-rankers into relevance judgment generators and integrate \ourmodel with them. 

Third, we only show \ourmodel's high effectiveness in predicting two primary metrics (RR@10 and nDCG@10) used at TREC DL 19--22~\citep{craswell2022,2021craswell,craswell2020,craswell2019}.
It is worthwhile to consider other metrics at various cutoffs in future work, e.g., nDCG@20 and MAP@100.

Fourth, 
while \ourmodel exhibits a promising \ac{QPP} quality and can be used in scenarios where \ac{QPP} quality is prioritized and users have a higher tolerance level for latency, e.g., patent search or post analysis, it is worth improving \ourmodel's efficiency in future to widen its scope of applications.
We plan to investigate 
(i)~the use of multiple GPUs because judging each item in a ranked list is independent of each other, (ii)~distilling knowledge from \acp{LLM} to smaller language models~\citep{gu2023knowledge},
(iii)~compressing \acp{LLM} by using lower-bit (e.g., 2-bit) quantization~\citep{chee2023quip} or using low-rank factorization~\citep{xu2023tensorgpt}, and
(iv) proposing an adaptive sampling approach that selects only a subset of documents from a ranked list for LLM-based relevance prediction to optimize the trade-off between judgment cost and QPP performance.

\begin{acks}
    We thank our reviewers and associate editor for their constructive feedback and suggestions.

    This research was partially supported 
    by the China Scholarship Council (CSC) under grant number 202106220041,
    the Hybrid Intelligence Center, a 10-year program funded by the Dutch Ministry of Education, Culture and Science through the Netherlands Organisation for Scientific Research, \url{https://hybrid-intelligence-centre.nl},
    project LESSEN with project number NWA.1389.20.183 of the research program NWA ORC 2020/21, which is (partly) financed by the Dutch Research Council (NWO),
    project ROBUST with project number KICH3.LTP.20.006, which is (partly) financed by the Dutch Research Council (NWO) and the Dutch Ministry of Economic Affairs and Climate Policy (EZK) under the program LTP KIC 2020-2023,
    and
    the FINDHR (Fairness and Intersectional Non-Discrimination in Human Recommendation) project that received funding from the European Union’s Horizon Europe research and innovation program under grant agreement No 101070212.
    
    All content represents the opinion of the authors, which is not necessarily shared or endorsed by their respective employers and/or sponsors.
\end{acks}

\bibliographystyle{ACM-Reference-Format}
\bibliography{references}

\end{document}